\documentclass{ieeeaccess}

\IEEEoverridecommandlockouts
\usepackage{cite}
\usepackage{amsmath,amssymb,amsfonts}
\usepackage{graphicx}
\usepackage{xcolor}
\usepackage{hyperref}

\def\BibTeX{{\rm B\kern-.05em{\sc i\kern-.025em b}\kern-.08em
    T\kern-.1667em\lower.7ex\hbox{E}\kern-.125emX}}
    
\usepackage{subfigure}
\usepackage{graphicx}
\usepackage{chemist}
\usepackage{algorithm}
\usepackage{algpseudocode}
\usepackage{empheq,etoolbox}
\usepackage{array}
\usepackage{graphics}
\usepackage{amsmath,amsfonts,amssymb}
\usepackage{eurosym}
\usepackage{xcolor}
\usepackage{braket}

\usepackage{hyperref}
\usepackage[app]{overpic}
\usepackage{physics}
\usepackage{braket}
\usepackage{bm}
\usepackage{mathrsfs}
\usepackage{caption}

\newcommand\underrel[2]{\mathrel{\mathop{#2}\limits_{#1}}}

\begin{document}

\history{Date of publication xxxx 00, 0000, date of current version xxxx 00, 0000.}
\doi{10.1109/TQE.2020.DOI}

\title{On the Robustness of QPE to Compute Ground Properties of Many-Electron Systems}

\author{
\uppercase{Wassil Sennane}\authorrefmark{1}
\and
\uppercase{Jérémie Messud}\authorrefmark{2}
}

\address[1]{QUANTSOC, Paris, France (email: \href{mailto:direction@qcsocrd.eu}{direction@qcsocrd.eu})}
\address[2]{TotalEnergies, Paris La Défense, France (email: \href{mailto:jeremie.messud@totalenergies.com}{jeremie.messud@totalenergies.com})}

\begin{abstract}
We propose an analysis of the Quantum Phase Estimation (QPE) algorithm applied to many-electron systems by investigating its free parameters such as the time step, number of phase qubits, initial state preparation, number of measurement shots, and other parameters related to the unitary operators implementation.
A deep understanding of these parameters and their impact on QPE probability of success and precision of the results is important to pave the way towards more automation of QPE applied to predictive computational chemistry and material science. We here explicit a constructive method to set the QPE free parameters for ground energy estimation and ground state projection, gathering disseminated results from previous works, refining these results and developing new conditions for achieving target performance. We detail the impact of the QPE `blurring function', related to discretization effects, and propose a method to overcome corresponding pathologies. We finally demonstrate that, using the conditions gathered here, the complexity of the Trotterized version of QPE tends to depend mostly on physical system properties and weakly on the number of phase qubits.
Various numerical results illustrate the impact of QPE free parameters on success probability and discretization effects. The impact of Trotterization and other features on the precision of the results are illustrated by first numerical simulations on the \chemform{H_2} molecule, that allows us to derive useful insights.
\end{abstract}


\titlepgskip=-15pt

\maketitle

\onecolumn{}

\section{Introduction}

We are interested in solving the stationary Schr\"odinger equation for bound states  of many-fermion systems (indexed by a non-negative integer $j$):
%
\begin{equation}
H \ket{\psi_j} = E_j \ket{\psi_j}, 
\end{equation}
and more precisely in computing the ground state energy $E_0$ (denoted ground energy in the following to lighten the notations)
and ground state $\ket{\psi_0}$ of many-electron systems. 
Finding the exact solution with classical computing leads to a cost that is exponential in the system size $N_S$, i.e., the number of spin-orbitals when the state is expanded on an orbital basis
\cite{gao2024distributed}. 
Approximate classical computing methods with a polynomial cost in $N_S$ have been developed, such as truncated CI \cite{zgid2012truncated}, density functional theory (DFT) \cite{kohn1965self} or tensor networks \cite{anselme2024combining, jamet2025anderson}. However, with strongly correlated many-electron systems, these classical methods may not lead to sufficiently precise ground state properties. This hampers potential applications of industrial interest such as predictive computational chemistry and material science \cite{freitas2025fundamental, lyra2022deriving, 
greene2022modelling}.
A promise of quantum computing is to drastically reduce the cost of the computation of exact ground properties of many-electron systems.

Quantum phase estimation (QPE) stands as a cornerstone quantum algorithm \cite{nielsen2010quantum}, leveraging the properties of controlled unitaries and Quantum Fourier Transform (QFT). 
Its roots lie in the algorithm developed by Shor in 1994 for prime-number factorization \cite{shor1994algorithms}, with the potential to exponentially reduce the computational complexity compared to classical computing \cite{tippeconnic2025breaking5bitellipticcurve, lin2014shor, xiao2022distributed}.
In 1995, Kitaev introduced the general QPE algorithm \cite{kitaev1995} and, since then, QPE found applications in various fields, e.g. linear systems resolution \cite{HHL2009}.
References \cite{Abrams1999} (1999) and \cite{Travaglione_2001} (2001) were among the first to study an usage of QPE applied to the Schr\"odinger equation and many-electron systems, with the potential to exponentially reduce the complexity of the computation of exact ground state properties compared to classical computing. Various developments in that field have been made especially since 2014, see  \cite{nusran2014application, Cruz2019OptimizingQP, o2019quantum, Pezz2020QuantumPE, lin2022heisenberg, Sugisaki2023ProjectiveMQ, papadopoulos2024reductive, barbieri2024multiphase, hualde2024quantum, ino2024workflow, ku2025benchmarking, akiba2025gpu, tranter2025high, paul2025applications, gu2025validation, zhou2025assessing, sorathia2025quantum, shukla2025practicalquantumphaseestimation, scali2025purifiedphaseestimationsamples}. 

The application of QPE to many-electron systems implies quite specific features.
Its practical deployment hinges on various pillars, that are quite specific and different from Shor's algorithm for instance:
an initial state $\ket{\psi_{\rm init}}$ having sufficient overlap with the exact ground state, a unitary transformation that involves the exponentiation of an Hamiltonian $H$ together with a free parameter $t$ (that we will explicit later), a non-trivial readout strategy (number of shots $m_\epsilon$), and a number $N$ of phase qubits that allows us to precisely estimate an energy 
and avoid pathologies due to discretization effects.
Additionally, approximate implementations of the unitary transformations can affect the quality of QPE. For example, 
the common Trotterized implementation introduces additional free parameters that control the accuracy of the implementation, called Trotter steps $n$ \cite{suzuki_general_1991, rajagopal_generalization_1999,CHEHADE2026430,ronfaut2026numerical}.
Finally, in addition to the ground energy estimation, QPE applied to many-electron systems has the potential to output in some circumstances an approximation of the ground state  \cite{Travaglione_2001}, which might be further used to compute observables other than the energy.
Fig. \ref{fig:QPEalgo} gives an illustration of the QPE features and free parameters, and their `localization' in the QPE circuit. 
\Figure[t!](topskip=0pt, botskip=0pt, midskip=0pt){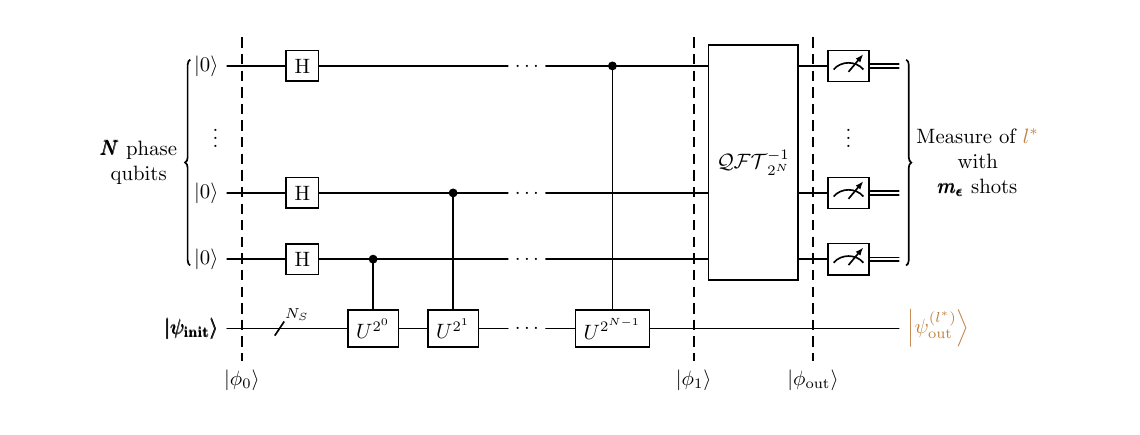}
{
QPE algorithm, with outputs $\left(\textcolor{brown}{l^*}, \textcolor{brown}{\ket{\psi_{\rm out}^{(l^*)}}}\right)$ and free parameters $\left(\pmb{t},\pmb{N},\pmb{\ket{\psi_{\rm init}}},\pmb{m_\epsilon},\pmb{n}\right)$ studied in this article. 
\label{fig:QPEalgo}
}

{A deep understanding of these parameters and their impact on the  probability of success and precision of the results is important to pave the way towards more automation of QPE,
i.e. to obtain a satisfying estimation of the ground energy $E_0$ and approximation of the ground state $\ket{\psi_0}$.
To our knowledge, while publications have dealt with QPE features related to some of these free parameters \cite{Travaglione_2001, shukla2025practicalquantumphaseestimation, ahmadi2010quantum,bauer2025postvariationalgroundstateestimation, childs_theory_2021}, 
a holistic method to coherently set these parameters has not been yet clearly stated.
We here explicit such a constructive method,
gathering disseminated results from previous works, refining these results and developing new conditions for achieving target performance.
We detail the impact of the QPE `blurring function', related to discretization effects, and propose a method to overcome corresponding pathologies.
We finally demonstrate that, using the conditions gathered here, the complexity of the Trotterized version of QPE tends to depend mostly on physical system properties and weakly on the number of phase qubits.
Various numerical results illustrate the impact of our points on the success probability and discretization effects.
The impact of Trotterization and other features on the precision of the results are illustrated by first numerical simulations on the \chemform{H_2} molecule, that allow us to derive useful insights.
}
This work aims to offer practical and standalone guidance towards more automation of QPE applied to predictive computational chemistry and material science.

The article is structured as follows: in Section \ref{sec:method} we provide an overview of QPE features, and in Section \ref{sec:summary2} we explain the challenge of defining the QPE free parameters and provide a summary of our proposals. The proofs and analysis of our findings are given in Sections \ref{sec:proofs1}-\ref{sec:trott}. 
Section \ref{sec:4} is dedicated to results on the \chemform{H_2} molecule.

\section{QPE Overview}\label{sec:method}

\subsection{Registers}

QPE uses two qubit registers \cite{shor1994algorithms, nielsen2010quantum}. In the case of an application to many-electron systems, these registers are:
\begin{itemize}
    \item
The phase qubit register $\ket{l}$ (also denoted by $\ket{k}$ in the following) whose output indirectly encodes an estimation of the eigenenergies of the system. The goal is to recover the ground energy $E_0$ from measurements on this register. It is a $N$-qubit register initialized as $\ket{0}=\ket{0}^{\otimes N}$, where $N$ represents an important free parameter controlling a trade-off between the quality of the result and the QPE resource need, as we will detail later.
    \item 
The system qubit register $\ket{\psi}$, whose output might represent a projection on the ground state $\ket{\psi_0}$ 
in some situations and thus might be used to compute additional observables other than the energy.
The number $N_S$ of qubits in this register is equal to the number of 
spin-orbitals required to describe the many-electron system.
It is initialized as $\ket{\psi_{\rm init}}$, which can be expressed using the basis of eigenstates of $H$:
\begin{equation}\label{eq:init}
\ket{\psi_{\rm init}} = \sum_{j\ge 0} c_j \ket{\psi_j}
\quad,\quad
\sum_{j\ge 0} |c_j|^2=1.
\end{equation}
$\ket{\psi_{\rm init}}$ represents another QPE free parameter important for the quality of QPE, as we will detail later.
In practice, it is obtained from a computation with polynomial cost, e.g. truncated CI  \cite{zgid2012truncated}, DFT \cite{kohn1965self}, tensor networks \cite{anselme2024combining, jamet2025anderson}; or even from parameterized quantum circuits \cite{fomichev2024initial,halder2021digitalquantumsimulationstrong}.
\end{itemize}
The complete QPE  state ($N+N_S$ qubits) is denoted by $\ket{\phi}$.

\subsection{Algorithm}

Fig. \ref{fig:QPEalgo} gives a visual overview of the QPE algorithm.
The phase register is first put in an equal superposition $\frac{1}{2^{N/2}}\sum_{k=0}^{2^{N}-1} \ket{k}$ by applying a Hadamard gate on each phase qubit.
Then, defining the following unitary operator that acts on the system register,
\begin{equation}\label{U}
  U = e^{-i2\pi H t},
\end{equation}
one implements a series of operators $U^{2^q}$ that are each controlled by the phase qubit $q \in \{0,...,N-1\}$.
The usage of Hartree (Ha) units for energy and atomic units for time is implicit in this formulation.
We have for the corresponding intermediate QPE state:
\begin{equation}\label{sumU}
  \ket{\phi_1}=\frac{1}{2^{N/2}} \sum_{k=0}^{2^N-1} \ket{k} U^k \ket{\psi_{\rm init}}
  =
  \frac{1}{2^{N/2}} \sum_{j\ge 0} c_j \sum_{k=0}^{2^N-1}
  e^{-i 2\pi E_j t  k}  \ket{k} \ket{\psi_j}
  =
  \frac{1}{2^{N/2}} \sum_{j\ge 0} c_j \sum_{k=0}^{2^N-1}
  e^{i 2\pi \theta_j^{(t)}  k}  \ket{k} \ket{\psi_j},
  \end{equation}
where (using the ceiling function):
\begin{align}\label{E_j}
\theta_j^{(t)} = -E_jt + \lceil E_jt \rceil = -E_jt \mod 1
\quad \in\quad [0,1[.
\end{align}

Lastly, an inverse 
QFT is performed on the phase register, which yields for the final QPE state just before phase register measurement:
\begin{equation}\label{finalstate}
  \ket{\phi_{out}} = \sum_{j\ge 0} \sum_{l=0}^{2^N-1} c_j\times
     f\left(\theta_j^{(t)}-\frac{l}{2^N}\right) \ket{l} \ket{\psi_j},
\end{equation}
where  \cite{Travaglione_2001, bauer2025postvariationalgroundstateestimation}:
\begin{align}\label{f}
    \forall \theta_j^{(t)}\in &~[0,1[,\quad \forall l\in\{0,...,2^N-1\}: \\
    &f\left(\theta_j^{(t)}-\frac{l}{2^N}\right) 
    = 
    \frac{1}{2^{N}}
    \frac{ 1-e^{i2\pi 2^N (\theta_j^{(t)}-\frac{l}{2^{N}})} }{1-e^{i2\pi(\theta_j^{(t)}-\frac{l}{2^{N}})}}
    = \frac{1}{2^{N}} \frac{\sin\left(\pi 2^N (\theta_j^{(t)}-\frac{l}{2^N})\right)}{\sin\left(\pi (\theta_j^{(t)}-\frac{l}{2^N})\right)}e^{i \pi (2^N - 1) (\theta_j^{(t)}-\frac{l}{2^{N}})},
    \nonumber
\end{align}
represents a `blurring function' related to discretization effects, as we will study below,
due to the fact that QPE approximates a continuous phase $\theta_j^{(t)}\in[0,1[$ by a discrete quantity $\frac{l}{2^N}$ \cite{Travaglione_2001}.
We have \cite{Travaglione_2001}:
\begin{align}\label{f2}
    \left|f\left(\theta_j^{(t)}-\frac{l}{2^N}\right)
    \right|^2
    = \frac{1}{2^{2N}} \frac{\sin^2\left(\pi 2^N \left(\theta_j^{(t)} - \frac{l}{2^{N}}\right)\right)}{\sin^2\left(\pi \left(\theta_j^{(t)} - \frac{l}{2^{N}}\right)\right)}
    \quad\ge\quad
     \frac{\sin^2\left(\pi 2^N \left(\theta_j^{(t)} - \frac{l}{2^{N}}\right)\right)}{\left(\pi 2^N \left(\theta_j^{(t)} - \frac{l}{2^{N}}\right)\right)^2}.
\end{align}
Denoting by $P(l)$ the probability that a measure of the phase register gives a value $l$,
\begin{equation}\label{Prob}
 P(l) = \sum_{j\ge 0} \lvert (\bra{\psi_j} \bra{l})\ket{\phi_{out}} \rvert^2
      = \sum_{j\ge 0} |c_j|^2 \left|f\left(\theta_j^{(t)}-\frac{l}{2^N}\right) \right|^2,
\end{equation}
we have:
\begin{align}\label{Prob-sum}
    \sum_{l=0}^{2^{N}-1} P(l) = 1,
\end{align}
which follows from two stronger relations, i.e.  \eqref{eq:init} (right) and
\footnote{
The relation $\frac{1}{2^{2N}} \sum_{l=0}^{2^{N}-1} \frac{\sin^2(\pi 2^N (x - \frac{l}{2^{N}}))}{\sin^2(\pi (x - \frac{l}{2^{N}}))}=1$ can be demonstrated using the infinite series $\frac{1}{\sin^2(\pi 2^N (x - \frac{l}{2^{N}}))}=\frac{1}{\pi^2}\sum_{k=-\infty}^{+\infty}\frac{1}{(2^N (x - \frac{l}{2^{N}})-k)^2}$.
}:
\begin{align}
    \sum_{l=0}^{2^{N}-1} 
    \left|f\left(\theta_j^{(t)}-\frac{l}{2^N}\right)\right|^2
    =1.
\label{eq:f^2}
\end{align}
$P(l)$-related features represent the main driver of QPE efficiency, as we will detail later.
From ~\eqref{finalstate}, we deduce that a measure of the phase register that gives a value $l$ projects the system register on the state:
\begin{equation}\label{ampj}
\ket{\psi^{(l)}_{out}} =
\sum_{j\ge 0} c_j^{(l)} \ket{\psi_j}
\quad,\quad
c_j^{(l)} = c_j \frac{f\left(\theta_j^{(t)}-\frac{l}{2^N}\right)}{\sqrt{P(l)}}.
\end{equation}

\subsection{Main QPE `ingredients'}

As we can see from ~\eqref{E_j}, \eqref{f}, \eqref{Prob} and \eqref{ampj}, various ingredients can affect the quality of QPE:
\begin{itemize}
    \item
The chosen time step $t$ and number of phase qubits $N$, which must be chosen by the user.
    \item
The properties of the function $f(\theta_j^{(t)}-\frac{l}{2^N})$ and especially of the corresponding $l$-direction probability distribution $|f(\theta_j^{(t)}-\frac{l}{2^N})|^2$. Two cases can be distinguished.
        If there exists a value $\frac{l_j}{2^N}$ in $\{0,\frac{1}{2^N},\frac{2}{2^N}...,1-\frac{1}{2^N}\}$
        such that $\frac{l_j}{2^N}=\theta_j^{(t)}$, then $f(\theta_j^{(t)}-\frac{l}{2^N})=\delta_{l_j,l}$ \cite{Travaglione_2001}.
        In the general  case, such an equality cannot be achieved exactly as the $\frac{l}{2^N}$ take discrete values constrained by the user's choice for $N$, whereas the $\theta_j^{(t)}$ take real values set by features of the many-electron system and the $t$ parameter choice. 
        Then, $|f(\theta_j^{(t)}-\frac{l}{2^N})|^2$ accounts for discretization effects, and has a peak height $<1$ with some dispersion around the peak \cite{Travaglione_2001}. The peak points on:
        \begin{align}\label{f2mp}
           l^* =  l_0 = \arg\max_l\left|f\left(\theta_0^{(t)}-\frac{l}{2^N}\right)\right|^2,
        \end{align}
        $\frac{l_j}{2^N}$ representing the value in $\{0,\frac{1}{2^N},\frac{2}{2^N}...,1-\frac{1}{2^N}\}$ that is the closest to $\theta_j^{(t)}$, i.e. satisfying: 
        \begin{align}\label{eq:ljbound}
                l_j-\frac{1}{2} \leq 2^N\theta_j^{(t)} <  l_j + \frac{1}{2}
                \quad\Rightarrow\quad 
                \left| \theta_j^{(t)} - \frac{l_j}{2^N} \right| \leq \frac{1}{2^{N+1}}.
        \end{align}
    \item
The properties of $\ket{\psi_{\rm init}}$, i.e. the $c_j=\bra{\psi_j}\ket{\psi_{\rm init}}$ values and the corresponding $j$-direction probability distribution $|c_j|^2$.
It is often mentioned that $\lvert c_0 \rvert^2= \lvert \bra{\psi_0}\ket{\psi_{\rm init}} \rvert^2$ should be close to $1$ so that QPE is efficient \cite{nielsen2010quantum, o2019quantum}, which will be discussed and refined below. 
    \item 
Both $|f(\theta_j^{(t)}-\frac{l}{2^N})|^2$ and $|c_j|^2$ are combined into the probability $P(l)$, ~\eqref{Prob}, that a phase register measurement gives $l$. We denote by:
\begin{align}\label{l*}
l^*=\arg\max_l P(l),
\end{align}
the most probable QPE phase measurement value.
Because $P(l)$ mixes various $f(\theta_j^{(t)}-\frac{l}{2^N})$, it contains as many peaks as the different $\theta_j^{(t)}$ values associated with non-negligible $c_j$ coefficients, each peak having a height proportional to the corresponding $|c_j|^2$ and some dispersion related to $|f(\theta_j^{(t)}-\frac{l}{2^N})|^2$ (discretization effects), which can `blur' the QPE phase measurement interpretation~
\footnote{
    In an ideal case where every 
    $\theta_j^{(t)} 2^N$ value is very close to an integer,
    we have $f(\theta_j^{(t)}-\frac{l}{2^N}) \approx \delta_{l_j,l}$ and ~\eqref{Prob} and~\eqref{ampj} simplify to
    $P(l) \approx \sum_{j\ge 0} |c_j|^2 \delta_{l_j,l}$ and $\ket{\psi^{(l)}_{out}} \propto \sum_{j\ge 0} c_j \delta_{l_j,l} \ket{\psi_j}$. This illustrates the `many peaks feature' even when the dispersion is negligible. 
}.
\end{itemize}

The importance of these ingredients implies that QPE applied to many-electron systems must not be considered as a black box. Indeed, corresponding free defined parameters can affect the quality of the QPE results, and a goal of this article is to detail how.

\subsection{Ground energy estimation and ground state projection}
\label{sec:gd_estim}

We now focus on the ground properties ($j=0$ in previous equations). As $l_0$ is related to the best estimation of the ground energy achievable with $N$ phase qubits, we would like to recover it from a series of QPE measures and compute, using  \eqref{E_j}:
\begin{align}\label{Eestim}
E_0= -\frac{\theta_0^{(t)}}{t} + \frac{1}{t}\lceil E_0t \rceil
\approx -\frac{1}{t}\frac{l_0}{2^N} + \frac{1}{t}\lceil E_0t \rceil.
\end{align}
This requires to know $\lceil E_0 t\rceil$, a challenge that can be addressed by an appropriate choice of $t$ as we will justify further. Recovering $l_0$ directly from measurements is possible for instance  when:
\begin{align}\label{l*=l0}
l^*=l_0\text{ (to be satisfied)}.
\end{align}
Then, reading the most probable phase directly gives $l_0$.
From~\eqref{Prob}, we deduce that~\eqref{l*=l0} can be satisfied if $P(l_0)>\max_{j\ne l_0}P(j)$, or equivalently if:
\begin{align}\label{proj_constr_delta}
\Delta_{(l_0)}>0,
\end{align}
%
where:
\begin{align}\label{proj_constr_delta2}
\Delta_{(l)}=P(l)-\max_{j\ne l}P(j).
\end{align}
Then, the QPE system register output state after a phase measurement that gives $l^*$ is, using the notations in ~\eqref{ampj}: 
\begin{align}
\ket{\psi^{(l_0)}_{out}} =
c_0^{(l_0)} \ket{\psi_0} + \sum_{j\ge 1} c_j^{(l_0)} \ket{\psi_j},
\end{align}
which can be used as a ground state $\ket{\psi_0}$ approximation (e.g. to compute observables other than the energy) 
if:
\begin{align}\label{proj_constr_2}
\left|c_0^{(l_0)}\right|^2\gg\left|c_j^{(l_0)}\right|^2 \quad\text{for any $j\ge 1$}.
\end{align}
In a weaker sense and in the spirit of  \cite{Travaglione_2001}, QPE can be considered as a step towards ground state projection if:
\begin{equation}\label{proj_constr_8}
\left|c_0^{(l_0)}\right|^2 
> \left|c_0\right|^2
.
\end{equation}

Another possibility is to `relax' the conditions derived in~\eqref{proj_constr_delta}-\eqref{proj_constr_delta2} and
consider carefully-chosen $e$-accurate measured $l$ values~\cite{nielsen2010quantum}, satisfying:
\begin{align}\label{e}
    |l - l_0| \leq e\quad,\quad\text{$e$ non-negative integer},
\end{align}
which represents a confidence interval associated to a higher confidence level $P(|l - l_0| \leq e) = \sum_{l = l_0-e}^{l_0+e}P(l) > P(l_0)$, as we will detail later.
Note that if such a procedure can improve the phase measurement efficiency, it can introduce challenges:
\begin{itemize}
    \item The 
    evaluation of an approximation of $l_0$ from a set of QPE measures (with controlled accuracy) for an usage in  \eqref{Eestim},
    \item The effectiveness of the ground state projection after a phase measurement $l$ close to the retained approximation $l_0$, fostering the conditions in ~\eqref{proj_constr_2}-\eqref{proj_constr_8}.
\end{itemize}
In the following, we will consider first the condition in ~\eqref{proj_constr_delta}-\eqref{proj_constr_delta2} and then detail aspects of the procedure related to  \eqref{e}.

Finally, we highlight that different $j$ values might satisfy $\theta_j^{(t)}=\theta_0^{(t)}$, not only in the case of ground state degeneracy but also beyond this case.
This can be due to the mapping of all eigenvalues $E_j$ into the $[0,1[$ interval within QPE, that can lead to eigen-phases `folding' due to the modulo $1$ in (\ref{E_j}), and to the discretization interval fostering potential overlap \cite{Travaglione_2001}.
We call the whole effect `QPE phase degeneracy'. 
This effect might be positive in terms of phase register measurement, reinforcing the probability associated to the ground phase. However, a condition like in ~\eqref{proj_constr_delta} should still be satisfied to ensure a good interpretation of the results,
and QPE phase degeneracy 
can be problematic as it can also reinforce the probabilities associated to `excited phases' \cite{Travaglione_2001}.
In terms of ground state projection, a QPE phase degeneracy related only to ground state degeneracy would output a combination in the subspace of degenerate ground states. This is acceptable, but the combination would depend on features of $\ket{\psi_{\rm init}}$ that are difficult to control. 

\section{Problem statement and summary of the results (constructive conditions on QPE free parameters)}
\label{sec:summary2}
We highlighted how QPE features and free parameters can affect the quality of the results {in terms of success probability and precision}: the time step $t$, the number of phase qubits $N$ (among others related to discretization effects), the initial state $\ket{\psi_{\rm init}}$ and the number of QPE shots. Approximate implementations of the  unitaries $U^{2^q}$ can also affect QPE, and the common Trotterized implementation precision depends on an additional free parameter called Trotter step.

The problem we will study in this article is: how to constructively define these free parameters to obtain a chemically precise estimation of the ground energy $E_0$ and a good approximation of the ground state $\ket{\psi_0}$ 
from $\ket{\psi^{(l^*)}_{out}}$,
{together with a good  success probability and a mitigation of discretization effects?
To our knowledge, no extensive study combining all aspects exists in the literature. 
We here explicit such a constructive method,
gathering disseminated results from previous works, e.g.  \cite{Travaglione_2001, ahmadi2010quantum,bauer2025postvariationalgroundstateestimation, childs_theory_2021,shukla2025practicalquantumphaseestimation}, refining these results and developing new conditions.}

Here is a summary of the main insights that will be derived and detailed in the next sections:
\begin{itemize}
    \item \textbf{$t$ conditions (section \ref{sec:time}).} 
    The first element to clarify when using QPE for molecular systems is the ability to recover the ground energy $E_0$ from the phase $\theta_0^{(t)}$, ~\eqref{Eestim}. This is non trivial because the required  $\lceil E_0 t\rceil$ is not known a priori. Our goal is to set a $t$ value such that $\lceil E_0t\rceil$ can be evaluated leveraging ceiling properties, which represents a gap in the literature to our knowledge.
    Our considerations will, among others, be based on:
    \begin{align}\label{eq:Delta_E}
    \Delta E=E_0-E_{\rm init}
    , \quad
    E_{\rm init}=\bra{\psi_{\rm init}}H\ket{\psi_{\rm init}},
    \end{align}
    recognizing that the known initial state energy $E_{\rm init}$ represents an approximation of $E_0$ whose accuracy can be defined through $\Delta E$.
    Three cases can be considered, the first two requiring prior knowledge:
    \begin{itemize}
        \item The first case, which occurs very rarely, supposes that we precisely know the order of magnitude of $\Delta E$ (through some experiment, database, etc.). This means that we know the smallest positive or negative integer $d$ such that $\Delta E\le-10^{-(d+1)}$. Then, 
        taking $t=10^{d}$ allows us to consider $\lceil E_0t\rceil=\lceil E_{\rm init}t\rceil$. 
        %
        \item The second case, which is more realistic, is suitable for a variety of situations where the inaccuracy $\frac{\Delta E}{E_{\rm init}}$ is reasonably large (in [33\%,100\%[).
        We then can take:
        \begin{align}
            t = -\frac{\alpha}{ E_{\rm init} }
            \quad,\quad
            \alpha \in \left[1,\frac{3}{2}\right]
            \quad\Rightarrow\quad
            \lceil E_0 t \rceil=\lceil E_{\rm init} t \rceil=-1.
        \end{align}
        %
        \item Given that the problem Hamiltonian is naturally expressed as a Linear Combination of Unitaries (LCU),
        \begin{align}\label{eq:LCU}
        H=\sum_\beta \gamma_\beta H_\beta,
        \end{align}
        where $H_\beta$ are unitaries defined from tensor products of Pauli operators acting on the system qubits,
        the third case considers:
        \begin{align}
            t=\frac{\alpha}{\sum_\beta |\gamma_\beta|}
            \quad,\quad \alpha\in\left[\frac{1}{2},1\right]
            \quad\Rightarrow\quad
            \lceil E_0 t \rceil=0.
        \end{align}
        This third case does not require any prior knowledge on the energy or the initial state and avoids eigen-phases folding due to the modulo $1$ in (\ref{E_j}) (a component of QPE phase degeneracy), making it the most common option. It usually leads to the smallest $t$ value, which is not a problem (e.g., we will demonstrate that our conditions lead to an independence on $t$ of the required resources for the Trotterized version of QPE).
        \end{itemize}
    {Note that in all cases $t$ tends to diminish as the size $N_S$ of the system increases.}
    \item \textbf{$N$ condition (section \ref{sec:phase}).} Once $t$ defined and the ambiguity on $\lceil E_0 t \rceil$ removed, we can set accordingly the number of phase qubits. Using QPE, we expect the ground (or any other) energy estimation to reach the chemical precision,
    \begin{align}
    \varepsilon_{\rm ch}=1.6\times10^{-3} \text{ Ha}.
    \end{align}
    For that, we will demonstrate the following condition must be satisfied:
    \begin{align}
    N \geq N_{\rm min}(t)=\left\lceil\log_2{\left(\frac{1}{t\varepsilon_{\rm ch}}\right)}\right\rceil -1.
    \end{align}
    This condition allows us to refine other proposals, e.g. the one in ~\cite{bauer2025postvariationalgroundstateestimation}.
    Interestingly, the minimum number of phase qubits $N_{\rm min}(t)$ is directly related to the choice made for $t$, a larger $t$ leading to a smaller $N_{\rm min}(t)$. 
    \item {\textbf{Mitigating discretization effects (sections \ref{sec:f}, \ref{sec:morephasequbits}  and \ref{sec:phase2}).}}
    {We will detail how discretization effects related to the behavior of $|f(\theta_j^{(t)}-\frac{l}{2^N})|^2$ depend non-trivially and non-regularly on $N$ (no convergence as $N$ increases), and can significantly impact the QPE probability $P(l)$, thus also the requirements on the initial state $\ket{\psi_{\rm init}}$ and the number of shots $m_\epsilon$ (next two bullet points).
    References \cite{Travaglione_2001, bauer2025postvariationalgroundstateestimation} commented the behavior of the $|f(\cdot)|^2$ distribution w.r.t. the continuous variable $\theta_j^{(t)}$, but it is the behavior w.r.t. the discrete variable $l$ which represents the QPE quality driver we will focus on.}
    Additionally, to avoid potential problems related to discretization effects, we will discuss how considering:
    \begin{align}
    N = N_{\rm min}(t) +a
    \quad,\quad\text{$a$ positive integer},
    \end{align}
    qubits could be interesting to facilitate QPE ground energy estimation, as taking:
    \begin{align}\label{everyfirst}
        e=2^{a-1}-1
    \end{align}
    ensures that any of the $2e+1$ corresponding values of $l$ that satisfy  \eqref{e}  (i.e. that are $e$-accurate w.r.t. $l_0$) lead to ground energy estimates that are closer to $E_0$ than the chemical precision.
    This can be leveraged in order to increase the probability of measuring chemically precise phases $P(|l - l_0| \leq e) = \sum_{l = l_0-e}^{l_0+e}P(l)$ or, in other terms, to increase the confidence level related to the chemical precise confidence interval $|l - l_0| \leq e$.
    But this comes with the challenges mentioned in section \ref{sec:gd_estim}. We suggest a complementary and constructive method which overcomes these limitations and helps for further ground state projection: perform a reduced number of shots of QPE circuits with $N = N_{\rm min}(t)+a$ phase qubits, for each $a \in \{0, 1,2,3\}$, and keep for the rest of the process the $a$ value that provides the most efficient $l^*$ (i.e. the case that provides the lowest variance around $l^*$, together with a high probability associated to the a window (or interval) of size $2e+1$).
    \item \textbf{$\ket{\psi_{\rm init}}$ condition (section \ref{sec:init}).} We analyze in detail the behaviors of $|f(\theta_j^{(t)}-\frac{l}{2^N})|^2$ and the probability $P(l)$ in the integer $l$-direction, going further than the works in  \cite{Travaglione_2001, bauer2025postvariationalgroundstateestimation}.
    While not summarized here, we highlight the fundamental conditions on $\ket{\psi_{\rm init}}$ that must be satisfied for $l^*$ to be equal to $l_0$, ~\eqref{l*=l0}, and for subsequent ground state projection following the $l^*$ measurement.
    Finally, we propose `average' and approximate conditions, that can be more or less loose according to the case but have the virtue for providing an idea of the required `quality' of the QPE initial state $\ket{\psi_{\rm init}}$.
    The first condition is related to ground energy estimation:
    \begin{align}\label{condqpe1}
    |c_0|^2 \gtrsim 0.6 \quad\text{on average w.r.t. $N$},
    \end{align}
    and the second to ground state projection: 
    \begin{align}\label{condqpe2}
    |c_0|^2\gtrsim \left|c_j\right|^2 0.6\times 5 \quad\text{for any $j\ge 1$}\quad\text{on average w.r.t. $N$}.
    \end{align}
    \item \textbf{Number of shots $m_\epsilon$ condition (section \ref{sec:shots}).} We build on the method of  \cite{shukla2025practicalquantumphaseestimation} and derive a different result that we believe is more adapted to QPE.
    An upper-bound for the minimum number of shots $m_\epsilon$ that ensures that the     higher count measured phase values are equal to $l^*$ {with probability $1-\epsilon$} is:
    \begin{equation}
         m_\epsilon = -\frac{2\ln \epsilon}{\Delta_{(l^*)}^2}
         \quad,\quad 0<\Delta_{(l^*)}\le 1,
    \end{equation}
    where $\Delta_{(l^*)}$ is defined through ~\eqref{proj_constr_delta2}.
    In practice, about a hundred QPE shots might be required to confirm the value of $l^*$, even with an initial state satisfying the previous conditions.
    Moreover, additional shots are required for ground state projection. Then, once the most probable phase $l^*$ is determined, the circuit is re-run until the measured phase equals $l^*$. When this is achieved (and the conditions described in this article met), the system register should be in the ground state of the Hamiltonian and can be further processed (to compute observables other than the energy).
    \item \textbf{Unitary approximation condition (number of Trotter steps illustration, section \ref{sec:trott}).} 
    All conditions developed above suppose that the implementations of the $U^{2^{q}}$ are exact. However, these implementations are in practice often approximate, denoted by $\mathcal{S}(U^{2^q})$.
    To foster that chemical precision can be reached, we justify the following first-order condition (using spectral norm): 
    \begin{equation}\label{Uineq}
        \left\lVert U^{2^q} - \mathcal{S}(U^{2^q}) \right\rVert \lesssim 2^q\frac{\pi}{2^{N_\text{min}(t)}}.
    \end{equation}
    Taking the example of the order-$p$ Trotterization approximation and inspired by the works of  \cite{suzuki_general_1991, mehendale_estimating_2025, childs_theory_2021},
    we demonstrate that an upper-bound for the minimum number of Trotter steps $n_\text{min}(q,t)$ compliant with  \eqref{Uineq}
    is given by:
    \begin{equation}
        n_\text{min}(q,t) 
        \approx\left\lceil \frac{2^q}{2^{N_\text{min}(t)}}\mathscr{C}_p\right\rceil
              \quad,\quad \mathscr{C}_p =
    \pi\left(\frac{C_p}{\varepsilon_{\rm ch}^{p+1}}\right)^\frac{1}{p},
    \end{equation}
where $C_p$ is obtained from commutators of the LCU components of $H$ and has the unit of an energy power $(p+1)$ \cite{childs_theory_2021,mehendale_estimating_2025}. Reference \cite{blunt_monte_2025} gives a methodology for estimating $C_p$ with a Monte-Carlo approach.
The overall number of QPE Trotter steps over the whole QPE circuit is:
\begin{align}
      n_\text{min-tot}\approx\left\lceil {2^a}\mathscr{C}_p \right\rceil.
\end{align}
Interestingly, $n_\text{min-tot}$ can be considered as independent of $N_\text{min}(t)$ and thus of $t$, and depends only on the number of phase qubits $a$ beyond the chemical precision. So, $n_\text{min-tot}$ tends to first-order to depend mostly on physical system features (especially on $N_S$) and weakly on the number of phase qubits, which is a property specific to the application of QPE to many-electron systems (the dependence of the Trotterization on the number of phase qubits that does not exceed an order of magnitude if $a \leq 3$). The complexity related to the Trotterized controlled-unitaries thus equals $\left\lceil {2^a}\mathscr{C}_p \right\rceil\times \mathscr{N}_p(N_S)$, where $\mathscr{N}_p(N_S)$ represents the complexity related to one order-$p$ Trotter step.
\end{itemize}

{
We conclude this summary by remarks on the complexity of QPE w.r.t. the size $N_S$ of the many-electron system, after our construction.
The implicit 
dependency of $t$ in $N_S$ (that should usually be polynomial) implies that $N_{\rm min}(t)$ tends to increase slowly when $N_S$ increases (the increase should usually be logarithmical). This means that the number of phase qubits usually does not represent a potential bottleneck for the method, contrary to Shor's algorithm applications. In practice, a few tens of phase qubits should be sufficient to reach the chemical precision for molecules with a few hundreds of electrons.
The complexity is thus mostly driven by the implementation of the controlled unitaries. For an order-$p$ Trotterization implementation, this complexity is dominated by $\mathscr{N}_p(N_S)$ which represents the cost of a single Trotter step - typically measured in terms of number of non-Clifford gates. This cost scales polynomially in $N_S$  \cite{PRXQuantum.4.020323}.
The total complexity is obtained multiplying $\left\lceil {2^a}\mathscr{C}_p \right\rceil$ by the necessary number of shots $m_\epsilon$.
The challenges and cost of preparing the system qubit register $\ket{\psi_{\rm init}}$ is not accounted for in these complexity analyses.
}

\section{Conditions on time step and minimum number of phase qubits}
\label{sec:proofs1}

\subsection{Time step}
\label{sec:time}

We propose to first set the $t$ value such that $\lceil E_0 t\rceil$ takes a known value which, to our knowledge, has not been extensively studied in the literature. 
We can start from the initial state energy $E_{\rm init}$, ~\eqref{eq:Delta_E}, 
    which is known and can be considered as an approximation of $E_0$ whose accuracy is qualified through the $\Delta E$ defined in ~\eqref{eq:Delta_E}.
    $E_{\rm init}$ is usually obtained from variational methods of polynomial cost \cite{ayral2025classical, anselme2024combining, jamet2025anderson, sennane2023calculating}, leading to $\Delta E\le 0$, $\Delta E= 0$ meaning perfect accuracy (which is obviously unrealistic).
    $\frac{\Delta E}{E_{\rm init}}\ge 0$ represents another measure of the accuracy, that we call inaccuracy percentage in the following ($0$\% meaning perfect accuracy and larger values meaning larger inaccuracy).
    Our goal is to define a pertinent $t$ from some prior knowledge, e.g. on $\Delta E$ or $\frac{\Delta E}{E_{\rm init}}$,
    such that $\lceil E_0t\rceil$ becomes known.
    Three cases can be considered:
\begin{enumerate}
        \item In the first case, which occurs very rarely but is of conceptual interest, we suppose that we know the order of magnitude of $\Delta E$ (through some experiment, database, etc.), i.e. we know the smallest positive or negative integer $d$ such that $\Delta E\le-10^{-(d+1)}$. Then, 
        taking $t=10^{d}$ ensures $\lceil E_0t\rceil=\lceil E_{\rm init}t\rceil$ and is one of the largest $t$ ensuring this relation,
        %
        %
        leaving to QPE the computation of only the digits that must be improved in $E_{\rm init}$ to recover $E_0$.
        We denote this value by $t_{\rm max}$, which yields an upper-bound for $t$:
        \begin{align} t\in]0,t_{max}] \text{ with } \left\{
        \begin{array}{l}
            t_{max}=10^d \\ \\
            d \text{ the smallest integer such that } \Delta E\le-10^{-(d+1)}
        \end{array} \right.
        \quad\Rightarrow\quad
        \lceil E_0 t \rceil=\lceil E_{\rm init} t \rceil. &
        \end{align}
        \item In the second case, which is more realistic, we suppose we 
        have much less precise information on $\Delta E$ and just have an idea of the value that bounds the inaccuracy $\frac{\Delta E}{E_{\rm init}}$, e.g. $\frac{1}{3}$ (meaning $E_{\rm init}$ is $\approx 33$\% inaccurate). We parameterize:
        \begin{align}
        t = -\frac{\alpha}{ E_{\rm init} }
        \quad,\quad \text{$\alpha>0$},
        \end{align}
        which leads to:
        \begin{align}
            \lceil E_0t \rceil = 
            \lceil (E_{\rm init}+\Delta E) t \rceil
            =\left\lceil -\alpha\left(1 + \frac{\Delta E}{E_{\rm init}}\right) \right\rceil
            ,\quad
            \lceil E_{\rm init} t \rceil = \left\lceil -\alpha \right\rceil.
        \end{align}
        Considering
        $\lceil -\alpha \rceil - 1 <  -\alpha\left(1 + \frac{\Delta E}{E_{\rm init}}\right) \le \lceil -\alpha \rceil$ leads to:
        \begin{align}
        \frac{\Delta E}{E_{\rm init}}<\frac{1-\lceil -\alpha \rceil}{\alpha}-1
        \quad\Rightarrow\quad
        \lceil E_0 t \rceil=\lceil E_{\rm init} t \rceil=\lceil -\alpha \rceil.
        \end{align}
        This relation allows us to relate the choice for $\alpha$ to our prior knowledge on the inaccuracy $\frac{\Delta E}{E_{\rm init}}$:
        \begin{itemize}
            \item  $\alpha=\frac{3}{2}$ $\Rightarrow$ $\lceil E_{\rm init} t \rceil=-1$ and $\frac{1-\lceil -\alpha \rceil}{\alpha}-1=\frac{1}{3}$, meaning             
            a $33$\% maximum inaccuracy.
            \item  If we consider $\alpha=1$ $\Rightarrow$ $\lceil E_{\rm init} t \rceil=-1$ (still) and $\frac{1-\lceil -\alpha \rceil}{\alpha}-1=1$, meaning   
            a $<100$\% maximum inaccuracy. 
            \item  Because of ceiling properties, the inaccuracy $\frac{1-\lceil -\alpha \rceil}{\alpha}-1$ does not necessarily have a regular behavior. So, in the following we consider:
            \begin{align}
            t = -\frac{\alpha}{ E_{\rm init} }
            \quad,\quad
            \alpha \in \left[1, \frac{3}{2}\right]
            \quad\Rightarrow\quad
            \lceil E_0 t \rceil=\lceil E_{\rm init} t \rceil=-1,
            \end{align}
        \end{itemize}
        which offer flexibility.
        Note that using $\alpha=\frac{3}{2}$ might be interesting. Indeed, it leads to:
        \begin{align}
            \theta_0^{(t)}=- E_0t + \lceil E_0t \rceil 
            = \alpha \frac{E_0}{E_{\rm init}} - 1 
            \underrel{\alpha=\frac{3}{2}}{=}
            \frac{3}{2}\left(1+\frac{\Delta E}{E_{\rm init}}\right)-1
            \underrel{\frac{\Delta E}{E_{\rm init}} \ll 1}{\rightarrow} \frac{1}{2},
        \end{align}
        and having $\theta_0^{(t)}$ close to $\frac{1}{2}$ might help to better leverage the information present in the first qubits of the phase register (compared to the case $\theta_0^{(t)}$ far from $\frac{1}{2}$ in which the first phase qubits tend to be all equal to $0$ or to $1$, depending on the case). This can somewhat contribute to the robustness of QPE.
        \item The third case does not require any prior knowledge on the energy or the initial state. It considers $t=\frac{\alpha}{||H||}$ where $||H||$ denotes the spectral norm of $H$ equal to $|E_0|$ and $\alpha\in]0,1[$, which leads to $\lceil E_0 t\rceil=\lceil -\alpha\rceil=0$. This is impractical because knowing the spectral norm of $H$ implies knowing the solution of the ground state problem. However, as the problem Hamiltonian is naturally expressed as a LCU \cite{nielsen2005fermionic, childs2012hamiltonian, loaiza2023reducing}, equation~\eqref{eq:LCU},
        we can easily compute the LCU coefficients 1-norm $\sum_\beta |\gamma_\beta|$ that satisfies $\sum_\beta |\gamma_\beta|\ge ||H||=|E_0|$ by triangular inequality; in practice, we can consider a strict inequality as the Hamiltonian LCU components are not collinear in practice. 
        We then have:
        \begin{align}\label{eq:LCU_t}
            t=\frac{\alpha}{\sum_\beta |\gamma_\beta|}< \frac{\alpha}{||H||}
            \quad,\quad \alpha\in]0,1]
            \quad\Rightarrow\quad
            \lceil E_0 t \rceil=\lceil E_j t \rceil=0.
        \end{align}
        Because $\theta_0^{(t)}=- E_0t + \lceil E_0t \rceil \le \alpha$, taking $\alpha \ge \frac{1}{2}$ is a necessary condition to have $\theta_0^{(t)}$ greater than $\frac{1}{2}$ and thus to leverage the information in the first qubits of the phase register.
        In practice, $\alpha=1$ is often taken, the LCU-coefficients 1-norm usually being a loose upper-bound on the spectral-norm.
        As \eqref{eq:LCU_t} is valid for all bound states $j$, it avoids eigen-phases `folding' due to the modulo $1$ in (\ref{E_j}) (the largest eigen-phases being mapped to values close to $0$ and the ground phase being mapped to the largest value, which is always less than one). This option is the most commonly used.
\end{enumerate}

Finally, note that $t$ tends to diminish  as the size $N_S$ of the system increases. This is because $|\Delta E|$, $|E_{\rm init}|$ and $\sum_\beta |\gamma_\beta|$ tend to increase w.r.t. $N_S$. This leads to an implicit dependency of $t$ on $N_S$ which affects the optimum number of phase qubits 
as we will see in the following.

\subsection{Minimum number of phase qubits}
\label{sec:phase}

Once $t$ has been set by one of the three methods proposed above (most often using the method 3 above), we can define the minimum number of phase qubits for a satisfying energy reconstruction.
Using QPE, we expect any energy estimation (thus also the estimation of the ground energy) to reach chemical precision $\varepsilon_{\rm ch}=1.6\times10^{-3}$ Ha. Using ~\eqref{E_j}, \eqref{eq:ljbound} and \eqref{Eestim} implies that for any $j\ge 0$:

\begin{align}\label{eq:N_constr}
        \left| \frac{1}{t}\frac{l_j}{2^N} - \frac{\theta_j^{(t)}}{t} \right|
        \leq \frac{1}{2^{N+1}t}\leq \varepsilon_{\rm ch}
        \quad\Rightarrow\quad N \geq N_{\rm min}(t)=\left\lceil\log_2{\left(\frac{1}{t\varepsilon_{\rm ch}}\right)}\right\rceil -1,
\end{align} 
where $N_{\rm min}(t)$ represents the minimum number of phase qubits to reach chemical precision.

Note that ~\eqref{eq:N_constr} allows us to refine other proposals, e.g. the one in  \cite{bauer2025postvariationalgroundstateestimation} that considers the energy gap instead of the chemical precision and where the $t$-dependency is implicit.
Also, choosing a larger $t$ allows us to reduce $N_{\rm min}(t)$. For instance, $N=20$ phase qubits are sufficient if $t=10^{-3}$ is pertinent; $N=10$ phase qubits are sufficient if $t=1$ is pertinent.

{
The implicit dependency of $t$ in $N_S$  (that should usually be polynomial) implies that $N_{\rm min}(t)$ tends to increase slowly when $N_S$ increases (the increase should usually be logarithmic). Fig. 4 in  \cite{bauer2025postvariationalgroundstateestimation} confirms this point even if the setting is slightly different as already mentioned.
In practice, few tens of phase qubits should be sufficient for molecules with few hundreds of electrons.
}

Using at least $N_{\rm min}(t)$ phase qubits ensures that any energy $E_j$ estimate obtained through $\frac{l_j}{2^N}$ reaches chemical precision.
However, taking more phase qubits such as
\begin{align}\label{eq:N_constr_more}
    N = N_{\rm min}(t) + a
    \quad,\quad\text{for a positive integer $a$},
\end{align}
can be interesting to foster the success of QPE,
as we will detail later. 

\section{Probability distributions, phase qubits, and conditions on initial state and number of shots}
\label{sec:proofs2}

\subsection{Analysis of $|f(.)|^2$}
\label{sec:f}

Let us illustrate the impact of the number $N$ of phase qubits on the integer $l$-direction probability distribution $|f(\theta_j^{(t)}-\frac{l}{2^N})|^2$.
References \cite{Travaglione_2001, bauer2025postvariationalgroundstateestimation} commented the behavior of the distribution w.r.t. the continuous variable $\theta_j^{(t)}$, but it is the behavior w.r.t. $l$ for a set of given real $\theta_j^{(t)}$ determined by the considered many-electron system which represents an important driver of the QPE quality. In the following, we focus on this behavior, which to our knowledge has not been extensively studied in the literature applied to many-electron systems.

\begin{figure}[H]
    \centering
    \includegraphics[width=\linewidth]{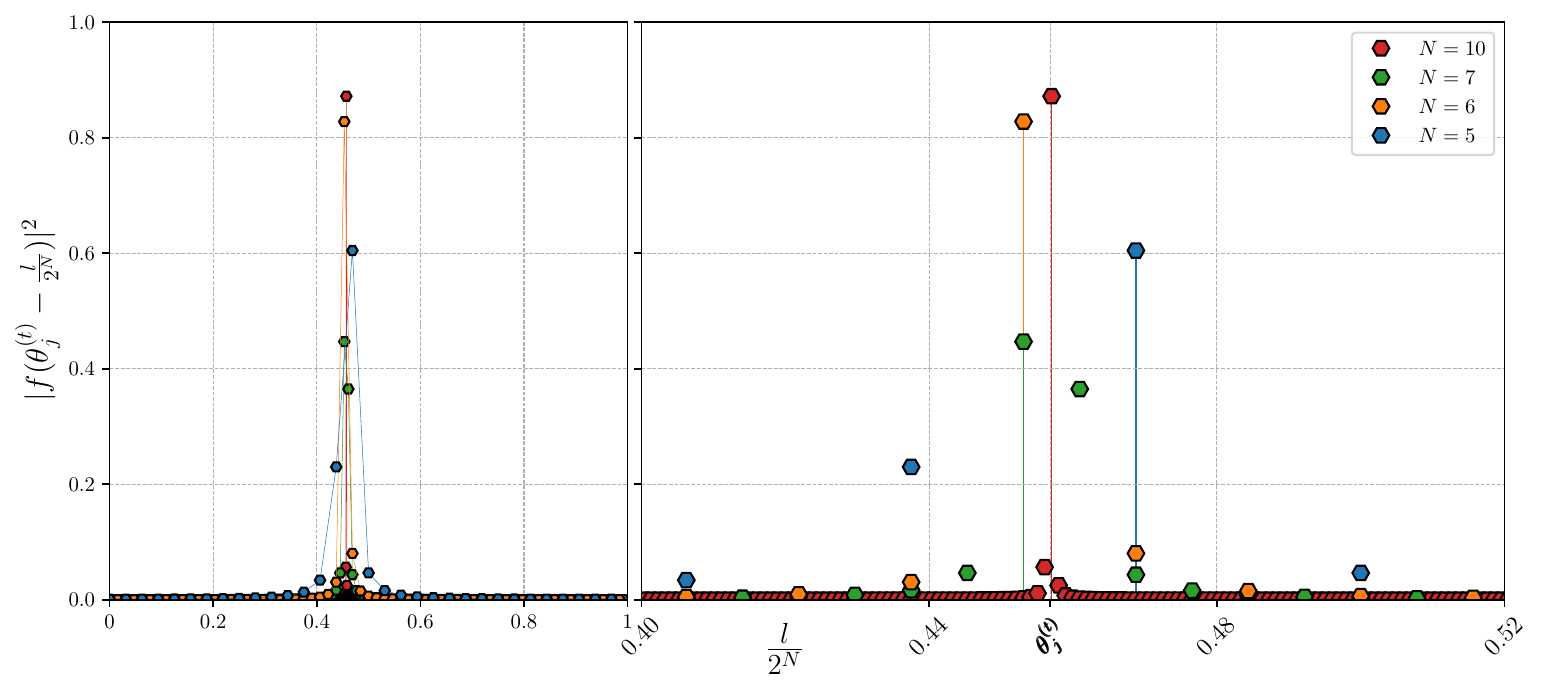}
    \put(-400, 228){(a)}
    \put(-165, 228){(b)}
    \caption{\textbf{(a)} $\lvert f(\theta_j^{(t)}-\frac{l}{2^N}) \rvert ^2$ as a function of the discrete variable $\frac{l}{2^N}$ with a fixed value of $\theta_j^{(t)}$, for several values of $N$.   \textbf{(b)}~Zoom of the left figure around $\theta_j^{(t)}$. Vertical lines denotes the $\frac{l_j^{(N)}}{2^N}$ value (with highest probability $\lvert f(\theta_j^{(t)}-\frac{l_j^{(N)}}{2^N}) \rvert^2$).}
    \label{fplot_v2}
\end{figure}

From ~\eqref{eq:ljbound}, we know that the value $\frac{l_j}{2^N}$ in $\{0,\frac{1}{2^N},\frac{2}{2^N}...,1-\frac{1}{2^N}\}$ that is the closest to a given $\theta_j^{(t)}$
is at-most $\frac{1}{2^{N+1}}$ far from $\theta_j^{(t)}$, implying:
\begin{align}\label{eq:ljbound2}
        \exists \kappa_j^{(N)}\in[0,1]:\quad\left| \theta_j^{(t)} - \frac{l_j^{(N)}}{2^N} \right| =\frac{\kappa_j^{(N)}}{2^{N+1}}.
\end{align}
It is important for the considerations in this section to highlight that the $l_j$ and $\kappa_j$ values are different for different $N$; this is why we write $l_j^{(N)}$ and $\kappa_j^{(N)}$ in the rest of this section and in the next section (only).
Equation~\eqref{eq:ljbound2} implies:
\begin{align}\label{lim1}
\lim_{N\rightarrow +\infty}\frac{l_j^{(N)}}{2^N}=\theta_j^{(t)},
\end{align}
which means we can approximate any $\theta_j^{(t)}$ by increasing $N$, the minimum $N$ to reach chemical precision on corresponding energies being  given by ~\eqref{eq:N_constr}. However, the maximum probabilities $|f(\theta_j^{(t)}-\frac{l_j^{(N)}}{2^N})|^2$ do not converge to a certain value as $N$ increases, which we explain now and can have an impact on the choice for $N$. Using ~\eqref{f2},~\eqref{eq:ljbound} and~\eqref{eq:ljbound2}, we have:
\begin{align}\label{f2bound}
    \left|f\left(\theta_j^{(t)}-\frac{l_j^{(N)}}{2^N}\right)\right|^2
    \,\ = \,\
    \frac{1}{2^{2N}} \frac{\sin^2\left(\frac{\pi\kappa_j^{(N)}}{2} \right)}{\sin^2\left(\frac{\pi\kappa_j^{(N)}}{2^{N+1}} \right)}
    \,\ \geq \,\
    \left(\frac{2}{\pi\kappa_j^{(N)}}\right)^2 \sin^2\left(\frac{\pi\kappa_j^{(N)}}{2} \right),
\end{align}
with the last term being reached asymptotically for reasonably large $N$ (here meaning $ N \gtrsim 5)$.
This leads to:
\begin{align}\label{f2values}
     \kappa_j^{(N)} & \rightarrow 1.0 & \Rightarrow  
     && \left\lvert f\left(\theta_j^{(t)}- \frac{l_j}{2^N}\right)\right\rvert^2 & {\rightarrow}  \,\ 0.41\\
     & \rightarrow 0.5 & 
     && &{\rightarrow}  \,\ 0.81 \nonumber\\
     & \rightarrow 0.0 &
     && &{\rightarrow} \,\ 1.00.\nonumber
\end{align}
Interestingly, 
the minimum probability related to the $\frac{l_j^{(N)}}{2^N}$ the closest to $\theta_j^{(t)}$ always satisfies \cite{Travaglione_2001}:
\begin{align}\label{min_prob_f}
\left|f\left(\theta_j^{(t)}-\frac{l_j^{(N)}}{2^N}\right)\right|^2{\ge} 0.41.
\end{align}
Note however that, in the general case, $|f(\theta_j^{(t)}-\frac{l_j^{(N)}}{2^N})|^2$ does not converge to $1$ as $N$ increases because $\kappa_j^{(N)}$ does not converge to $0$.
Indeed, $\kappa_j^{(N)}$ will tend to `oscillate' in $[0,1]$ as $N$ increases, which is due to the corresponding change in the discretization interval (e.g., for a specific $N$ value the discretization interval can be such that $\kappa_j^{(N)}$ is small; then, switching to $N+1$ phase qubits can lead to a higher value of $\kappa_j^{(N)}$; this will change again with $N+2$ phase qubits, etc.)
An important consequence is that $\lvert f(\theta_j^{(t)}-\frac{l_j^{(N)}}{2^N}) \rvert^2$ will tend to `oscillate' in $[0.41,1]$ as $N$ increases.
On average, we have:
\begin{align}\label{f2average}
    \kappa_j^{(N)}\approx 0.5\quad\text{on average \% $N$}
    \quad\Rightarrow\quad
    \left\lvert f\left(\theta_j^{(t)}- \frac{l_j^{(N)}}{2^N}\right)\right\rvert^2 \approx 0.81\quad\text{on average w.r.t. $N$}.
\end{align}
However, in the specific case where $2^N\theta_j^{(t)}$ is becomes very close to an integer for a given $N$ value, we have $\kappa_j^{(N)}\approx 0$ and $|f(\theta_j^{(t)}-\frac{l}{2^N})|^2\approx \delta_{l_j,l}$ for this $N$ value and any other greater $N$ value.
Fig. \ref{fplot_v2} illustrates these features. It shows for a given $\theta_j^{(t)}$ the behavior of $|f(\theta_j^{(t)}-\frac{l}{2^N})|^2$ as a function of $\frac{l}{2^N}$ for several values of $N$ between $5$ and $10$. 
We observe that $N=10$ allows us to obtain the $\frac{l_j^{(N)}}{2^N}$ value the closest to $\theta_j^{(t)}$, in agreement with the tendency ~\eqref{eq:ljbound2}. The associated probability $|f(\theta_j^{(t)}-\frac{l_j^{(N)}}{2^N})|^2$ is close to $0.87$ meaning a $\kappa_j^{(N=10)}\approx 0.3$.
Overall, $|\theta_j^{(t)}-\frac{l_j^{(N)}}{2^N}|$ tends to decrease as $N$ increases, but the decrease is not regular. For instance, we observe that $N=6$ and $N=7$ lead to almost the same $\frac{l_j}{2^N}$ values, for the reason related to the $\kappa_j^{(N)}$ behavior mentioned above. 
The associated probabilities $|f(\theta_j^{(t)}-\frac{l_j^{(N)}}{2^N})|^2$ do not evolve monotonically as $N$ increases: the probability associated to $N=6$ (close to 0.81 and related to $\kappa_j^{(N=6)}\approx 0.5$) is much higher than the one associated to $N=7$ (close to 0.45 and related to $\kappa_j^{(N=7)}\approx 0.95$).
Also, the two most probable values in the case $N=7$ are almost equiprobable, meaning that the real $\theta_j^{(t)}$ lies almost in-between a discretization interval ($\kappa_j^{(N=7)}\approx 0.95$ traducing this).
Choosing a sufficiently large $N$ is important to ensure the $\frac{l_j^{(N)}}{2^N}$ values become sufficiently close to $\theta_j^{(t)}$ (in practice that they reach the chemical precision on the corresponding energy). 
However, this does not ensure a large associated probability $\lvert f(\theta_j^{(t)}-\frac{l_j^{(N)}}{2^N}) \rvert ^2$ (i.e. a small associated $\kappa_j^{(N)}$). We observe that these probabilities are all between $0.41$ and $1$, in agreement with ~\eqref{f2values}, but their variation w.r.t. $N$ tends to 'oscillate' which makes it \textit{a priori} difficult to approach $1$ in a controlled way.

Considering a larger interval around $l_j^{(N)}$, as in ~\eqref{e},
\begin{align}\label{lj-e}
    \left|l - l_j^{(N)}\right| \leq e\quad,\quad\text{$e$ non-negative integer},
\end{align}
allows to increase the probability of success. The following relation can be proven :
\begin{align}\label{demo_e}
\sum_{l=l_j^{(N)}-e}^{l_j^{(N)}+e} \left|f\left(\theta_j^{(t)}-\frac{l}{2^N}\right)\right|^2
\quad \ge \quad
    \sum_{k=-e}^{e}
     \frac{\sin^2\left(\pi 2^N \left(\theta_j^{(t)} -  \frac{l_j^{(N)}}{2^{N}} \right)-\pi k\right)}{\pi^2\left(2^N \left(\theta_j^{(t)} -  \frac{l_j^{(N)}}{2^{N}}\right) - k\right)^2}
\quad \ge \quad
    \frac{1}{\pi^2}\sum_{k=-e}^{e}
     \frac{1}{\left(\frac{1}{2} - k\right)^2}
     ,
\end{align}
where we used ~\eqref{f2} for the first inequality and ~\eqref{eq:ljbound2} that implies $2^N|\theta_j^{(t)} - \frac{l_j^{(N)}}{2^{N}}|\le \frac{1}{2}$ for the second inequality
\footnote{
We can demonstrate that: $\forall \delta\in[-\frac{1}{2},\frac{1}{2}]: \sum_{k=-e}^{e}\frac{\sin^2\left(\pi (\delta- k)\right)}{\left(\pi(\delta - k)\right)^2}\ge \sum_{k=-e}^{e}\frac{\sin^2\left(\pi \frac{1}{2}-\pi k\right)}{\pi^2\left(\frac{1}{2} - k\right)^2}=\sum_{k=-e}^{e}\frac{1}{\pi^2\left(\frac{1}{2} - k\right)^2}$.
}. 
With $e=0$, we recover the value in  \eqref{min_prob_f}. With $e=1$ (meaning we sum the probabilities related to $l_j^{(N)}$ and to the two closest $l$ values), we have:
\begin{align}
\sum_{l=l_j^{(N)}-1}^{l_j^{(N)}+1} \left|f\left(\theta_j^{(t)}-\frac{l}{2^N}\right)\right|^2
\,\ &\ge \,\
0.85\\
&\approx \,\ 0.92\quad\text{on average w.r.t. $N$}.\nonumber
\end{align}
Fig. \ref{fplot_v2} confirms that the three most probable values sum to $\approx 0.86$ for $N=5$ and $N=7$, and beyond $0.92$ for $N=6$ and $N=10$, suggesting a non-regular behavior. We now start to investigate how this feature might be used when taking $N> N_{\rm min}(t)$, to leverage  more peaks around the ground phase than only the most probable one.

\subsection{Taking more phase qubits and \lowercase{$|f(.)|^2$} behavior}
\label{sec:morephasequbits}

Let us now investigate the case where $N= N_{\rm min}(t) + a$ with $a$ a positive integer, ~\eqref{eq:N_constr_more}. Given ~\eqref{eq:N_constr}, there can be more than one value of $l \in \{0, ..., 2^N-1\}$ that yields reconstructed ground energies within chemical precision. These values satisfy:
\begin{align}\label{eq:l_min_a_chem_acc-}
    \left\lvert\theta_j^{(t)} - \frac{l}{2^{N_{\rm min}(t)+a}}\right\rvert  \leq t\varepsilon_{\rm ch} = \frac{1}{2^{N_{\rm min}(t)+1}}
\quad\Rightarrow\quad
    \left\lvert 2^{N_{\rm min}(t)+a}\theta_j^{(t)} - l\right\rvert  \leq  2^{a-1}.
\end{align}
Our goal here is to define the number of peaks around the ground phase which satisfies the chemical precision constraint, i.e. the corresponding $e$ value in ~\eqref{lj-e}-\eqref{demo_e}, as a function of the $a$ value. From ~\eqref{eq:ljbound2}, we have:
\begin{align}\label{eq:l_min_a_case2}
     2^{N_{\rm min}(t)+a}\theta_j^{(t)} = l^{(N_{\rm min}(t)+a)}_j + sign\times \frac{\kappa_j^{(N_{\rm min}(t)+a)}}{2},
\end{align}
where $sign\in\{-1,1\}$ represents the sign of $(2^{N_{\rm min}(t)+a}\theta_j^{(t)} - l^{(N_{\rm min}(t)+a)}_j)$.
When  \eqref{eq:l_min_a_case2} is used in  \eqref{eq:l_min_a_chem_acc-}, we obtain:
\begin{align}\label{eq:e12}
   -e_1 \leq l - l^{(N_{\rm min}(t)+a)}_j \leq  e_2,
\end{align}
where:
\begin{align}
   e_1=\left\lfloor2^{a-1}-sign\times \frac{\kappa_j^{(N_{\rm min}(t)+a)}}{2} \right\rfloor
   \quad,\quad
   e_2=\left\lfloor2^{a-1} +sign\times \frac{\kappa_j^{(N_{\rm min}(t)+a)}}{2}\right\rfloor.
\end{align}
We have (remind that $\kappa_j^{(N_{\rm min}(t)+a)}\in[0,1]$):
\begin{align}
&\text{ if } a=0:&&&&e_1=e_2=0 
\quad\Rightarrow \quad l=l^{(N_{\rm min}(t)+a)}_j\\
&\text{ if } a\ge 1:&&\text{ if } \kappa_0^{(N_{\rm min}(t)+a)}=0:&&e_1=e_2=2^{a-1} \nonumber\\
&&&\text{ if } sign > 0:&&e_1=2^{a-1}-1 \text{ and } e_2=2^{a-1}  \nonumber\\
&&&\text{ if } sign < 0:&&e_1=2^{a-1} \text{ and } e_2=2^{a-1}-1.\nonumber
\end{align}
We note that, when $\kappa_j^{(N_{\rm min}(t)+a)}\ne 0$, there is always one `side' in  \eqref{eq:e12} with an `additional' $l$-value that satisfies the chemical precision condition, due to the fact that $2^{N_{\rm min}(t)+a}\theta_j^{(t)}$ is on the same side of $l^{(N_{\rm min}(t)+a)}_j$. 
A more restrictive but `symmetric' condition of the kind of  \eqref{lj-e} is obtained by taking $e=\min(e_1,e_2)$, which leads to:
\begin{align}\label{eq:le2}
    \left\lvert l - l^{(N_{\rm min}(t)+a)}_j\right\rvert \leq e=2^{a-1}-1 \quad\text{for}\quad a\ge 1.
\end{align}
Thus, we defined the $e$ value corresponding to the additional number of qubits $a$ to satisfy the chemical precision condition (the result is slightly different from the one in section 5.2.1 of  \cite{nielsen2010quantum} because of specific features of our framework). This allows to consider additional peaks into $|f(.)|^2$, ~\eqref{demo_e}, and thus allows for a smoother evolution of the corresponding probability.

\subsection{Analysis of $P(.)$, initial state conditions and effect of taking more phase qubits}
\label{sec:init}

We study the features of the main driver of the QPE efficiency, $P(l)$, ~\eqref{Prob}, which are driven by the input state-related probability distribution $|c_j|^2$ `filtered' by the $|f(\theta_j^{(t)}-\frac{l}{2^N}) |^2$. We will develop conditions on the quality of $\ket{\psi_{\rm init}}$, i.e. on $|c_0|^2$ which is related to the overlap between $\ket{\psi_{\rm init}}$ and $\ket{\psi_0}$, which represents another free parameter of QPE. For the remaining of this article we focus on the ground properties, $j=0$, and come back to a notation where the $N$ dependency is implicit. We have, using  \eqref{Prob}:
\begin{align}\label{Prob2}
 P(l_0) 
 \,\ =\,\ \sum_{j\ge 0} |c_j|^2 \left|f\left(\theta_j^{(t)}-\frac{l_0}{2^N}\right) \right|^2
 \,\ \ge\,\ |c_0|^2 \left|f\left(\theta_0^{(t)}-\frac{l_0}{2^N}\right) \right|^2&.
\end{align}

We first focus on ~\eqref{proj_constr_delta} that provides the condition so that the most probable phase measurement $l^*$ is associated to the best achievable estimation of the ground energy $l_0$, but this condition in impractical.
Considering $P(l_0)> 0.5$ represents a sufficient and more practical condition, leading to:
\begin{align}\label{proj_constr_00}
|c_0|^2
>
\frac{0.5}{\left|f\left(\theta_0^{(t)}-\frac{l_0}{2^N}\right)\right|^2}
- \sum_{j\ge 1} |c_{j}|^2 \Lambda_j
\quad,\quad
\Lambda_j=\frac{\left|f\left(\theta_j^{(t)}-\frac{l_0}{2^N}\right)\right|^2}{ \left|f\left(\theta_0^{(t)}-\frac{l_0}{2^N}\right)\right|^2}.
\end{align}

For ground state projection, the requirements in ~\eqref{proj_constr_2} or \eqref{proj_constr_8} (weaker condition) must be satisfied. After some calculation using ~\eqref{Prob} and~\eqref{ampj}, the first requirement leads to:
\begin{align}\label{ccc}
|c_0|^2\gg \left|c_j\right|^2\Lambda_j\quad\text{for any $j\ge 1$},
\end{align}
and the second (weaker) to:
\begin{align}\label{ddd}
\sum_{j\ge 1} |c_j|^2
\ge
\sum_{j\ge 1} |c_j|^2 \Lambda_j.
\end{align}

We see that the properties of the $\Lambda_j$ are driving both ground phase estimation, ~\eqref{proj_constr_00}, and ground state projection, ~\eqref{ccc}-\eqref{ddd} but differently: 
large $|c_j|^2 \Lambda_j$ values for $j\ge 1$ can be problematic for ground state projection but can help ground phase estimation, and conversely
\footnote{
A link can be done with the already mentioned properties of QPE phase degeneracy. 
}.
The behavior of $\Lambda_j$ w.r.t. $j$ is driven by the behavior of the $|f(\theta_j^{(t)}-\frac{l_0}{2^N})|^2$ w.r.t. the distribution of the $\theta_j^{(t)}$, at a given $N$ value.
It is shown in  \cite{Travaglione_2001} that, for sufficiently large $N$ (meaning $N\gtrsim 3$), we have:
\begin{align}\label{eq:Hset}
    \forall i \in \mathbb{H}_0=\left\{ i\in\{0,...,2^N-1\} \text{ such that: } \left| \theta_i^{(t)} - \frac{l_0}{2^N} \right| > \frac{1}{2^N} \right\}:&&\quad&
    \left|f\left(\theta_i^{(t)}-\frac{l_0}{2^N}\right)\right|^2\le 0.05,\\
    \forall i \in\overline{\mathbb{H}}_0\setminus l_0:&&\quad&
    \left|f\left(\theta_i^{(t)}-\frac{l_0}{2^N}\right)\right|^2\le 1.\nonumber
\end{align}
From ~\eqref{min_prob_f}, we deduce:
\begin{align}
\forall i \in\mathbb{H}_0: \Lambda_i\lesssim 0.12
\quad,\quad
\forall j \in\overline{\mathbb{H}}_0\setminus l_0: 
\Lambda_j\lesssim 2.4
\quad,\quad
\Lambda_j\approx 0.6 \quad\text{on average \% $j\in \overline{\mathbb{H}}_0\setminus l_0$ and $N$}.
\end{align}
The most problematic case for QPE ground state projection is when $\Lambda_j>1$, which occurs when the peak of $|f(\theta_j^{(t)}-\frac{l_0}{2^N})|^2$ related to an excited phase is higher than the peak of the ground phase (due to discretization effect and the distribution of the $\theta_j^{(t)}$). A sufficiently large $|c_0|^2$ together with the requirement that no large $|c_j|^2$ exist in $\ket{\psi_{\rm init}}$ for any $j\in\overline{\mathbb{H}}_0\setminus l_0$ should be sufficient for QPE to be used for ground energy estimation and state projection.
Reference \cite{Travaglione_2001} did an analysis in this direction, highlighting the delicate balance that this condition requires, which is difficult to control by the user.
More pragmatically, we propose the following `average' conditions, the first being sufficient for ~\eqref{proj_constr_00} to hold:
\begin{align}\label{condqpe1}
|c_0|^2 > \frac{0.5}{\left|f\left(\theta_0^{(t)}-\frac{l_0}{2^N}\right)\right|^2 }\approx 0.6 \quad\text{on average w.r.t. $N$},
\end{align}
and the second being intermediate between ~\eqref{ccc} and~\eqref{ddd}:
\begin{align}\label{condqpe2}
|c_0|^2\gtrsim \left|c_j\right|^2 0.6\times 5 \quad\text{for any $j\ge 1$}\quad\text{on average w.r.t. $N$}.
\end{align}
These requirements are approximate (and can be more or less loose according to the case) but provide an idea of the required quality (or purity) of the QPE initial state $\ket{\psi_{\rm init}}$.

\begin{figure}[H]
    \centering
    \includegraphics[width=\linewidth]{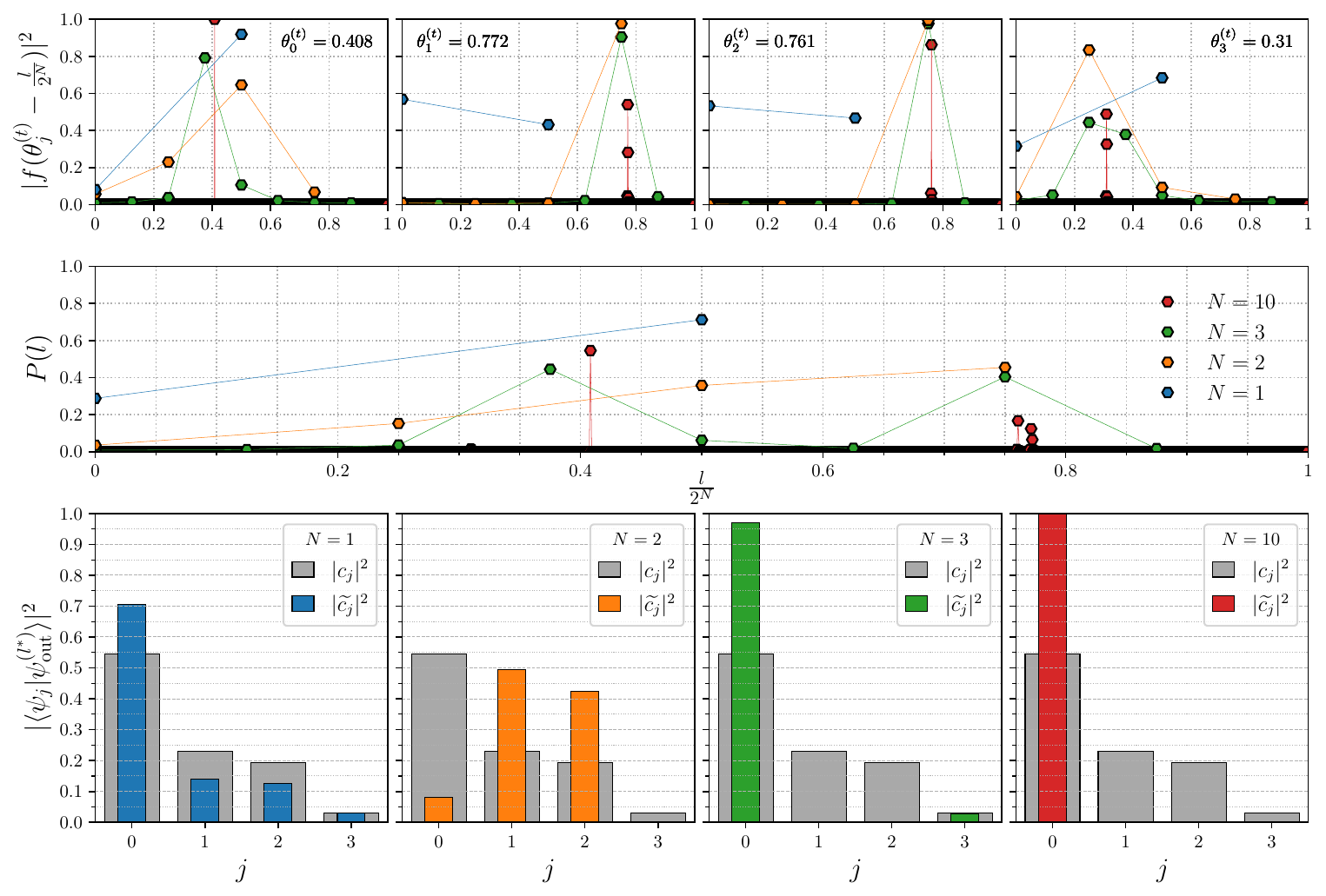}
    \put(-430, 345){(a)}
    \put(-310, 345){(b)}
    \put(-190, 345){(c)}
    \put(-70, 345){(d)}
    \put(-255, 245){(e)}
    \put(-430, 153){(f)}
    \put(-310, 153){(g)}
    \put(-190, 153){(h)}
    \put(-70, 153){(i)}
    \caption{System with $4$ eigenstates and randomly generated $\theta_j^{(t)}$. \textbf{(a), (b), (c), (d)} $\lvert f(\theta_j^{(t)}-\frac{l}{2^N}) \rvert^2$ as a function of $\frac{l}{2^N}$  for several $N$ values. \textbf{(e)}  $P(l)$ as a function of $\frac{l}{2^N}$ for several $N$ values. \textbf{(f), (g), (h), (i)} Overlap between each exact eigenstate of $H$ and: the initial state ($|c_j|^2$, grey), and the final state after phase measurement ($|c_j^{(l^*)}|^2$, color), for several  $N$ values.}
    \label{fig_f_discrete_1}
\end{figure}

Fig. \ref{fig_f_discrete_1} illustrates these QPE features. A system with $4$ eigenstates is considered, with random values of $\theta_j^{(t)}$ that are indicated in the \textbf{(a), (b), (c), (d)} panels. The $c_j$ have been generated to mimic a reasonable $\ket{\psi_{\rm init}}$ and satisfy the conditions in ~\eqref{condqpe1} and~\eqref{condqpe2}, the corresponding $|c_j|^2$ values being indicated in grey in the bottom figures.
We observe in Fig. \ref{fig_f_discrete_1} \textbf{(a), (b), (c), (d)} (similarly to Fig. \ref{fplot_v2}) how the $\lvert f(\theta_j^{(t)}-\frac{l}{2^N}) \rvert^2$ `filters' act, the cases $N=3$ and $10$ offering the best potential for an precise estimation of $\theta_0^{(t)}=0.408$. 
%
Fig. \ref{fig_f_discrete_1} \textbf{(e)} shows the corresponding $P(l)$ as a function of $\frac{l}{2^N}$. We can see that the cases $N=1,3$ and $10$ lead to a most probable phase $\frac{l^*}{2^N}$ that is the closest to $\theta_0^{(t)}=0.408$ available in the discretization interval ($N=10$ giving the most precise estimation as awaited). 
The case $N=3$ leads however to a most probable phase $\frac{l^*}{2^N}$ far from the ground phase $\theta_0^{(t)}$ and close to $\theta_1^{(t)}=0.772$ and $\theta_2^{(t)}=0.761$, due to a $\lvert f(\theta_j^{(t)}-\frac{l}{2^N}) \rvert^2$ close to one around $\theta_1^{(t)}$ and $\theta_2^{(t)}$, and non-negligible amplitudes $|c_1|^2$ and $|c_2|^2$.
This leads to a case with a quite small $|f(\theta_0^{(t)}-\frac{l_0}{2^N})|^2$ and quite large $\Lambda_j$ for $j=1-3$, so that ~\eqref{proj_constr_00}-\eqref{ddd} are not satisfied and our `average' conditions in ~\eqref{condqpe1}-\eqref{condqpe2} do not hold (we are far form the average here).

The bottom panels \textbf{(f), (g), (h), (i)} in Fig. \ref{fig_f_discrete_1} show the amplitude associated to each eigenvector of $H$ in the QPE system register: the initial QPE state in grey (related to the amplitudes in $\ket{\psi_{\rm init}}$), and the QPE state after a phase measurement that gives $l^*$ in color (related to the amplitudes in $\ket{\psi^{(l^*)}_{out}}$). 
The $N=1$ case represents a step towards ground state projection, and the $N=3$ and $10$ cases perform very good ground state projection. Indeed, in the latter cases, corresponding $\Lambda_1$ and $\Lambda_2$ are smaller,
and the ratio $\frac{|f(\theta_0^{(t)}-\frac{l_0}{2^N})|^2}{P(l_0)}$ is close to $1.7$. According to ~\eqref{ampj}, this leads to a massive amplification of the ground state. The amplification is even stronger in the $N=10$ case.
Interestingly, we notice how adding $1$ phase qubit and passing from $N=2$ to $N=3$ completely changes the situation according to behavior described in section \ref{sec:f}.

\begin{figure}[!h]
    \centering
    \includegraphics[width=0.75\linewidth]{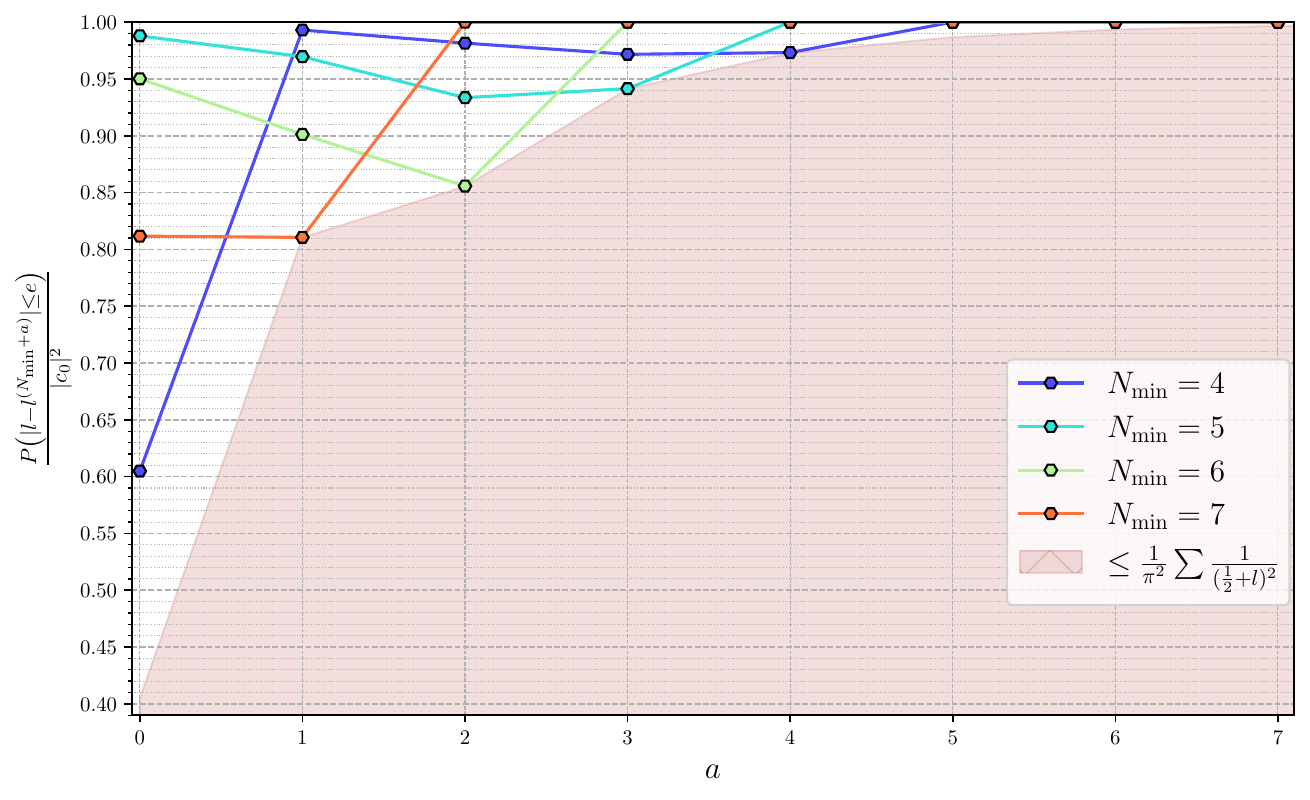}
    \caption{Probability to measure a value of $l_0$ satisfying the case defined by \eqref{e}, \eqref{eq:N_constr_more} and \eqref{eq:le2}, as a function of the $a$ qubits added to $N_{\rm min}$.
    The system 
    considers the same $\theta_j^{(t)}$ and $c_j$ than the case in Fig. \ref{fig_f_discrete_1}, but not the same numbers of phase qubits. The pink filled area represents the lower bound in \eqref{Prob2bis}.}
    \label{fig_P_relax_minor}
\end{figure}

We conclude this section by considering the case in ~\eqref{e}, resulting from the addition of $a$ qubits beyond the chemical precision, ~\eqref{eq:N_constr_more} and \eqref{eq:le2},
which can help to increase the QPE probability of success:
\begin{align}\label{Prob2bis}
    \sum_{l=l_0-e}^{l_0+e}  P(l)
    \ge
    |c_0|^2 \sum_{l=l_0-e}^{l_0+e} \left|f\left(\theta_0^{(t)}-\frac{l}{2^N}\right) \right|^2
     \ge 
    |c_0|^2\frac{1}{\pi^2}\sum_{k=-e}^{e}
     \frac{1}{\left(\frac{1}{2} - k\right)^2},
\end{align}
%
where the lower-bound in the right hand side is obtained using  \eqref{demo_e}.
Fig.~\ref{fig_P_relax_minor} represents the increased probability resulting from taking more phase qubits, for the same $\theta_j^{(t)}$ and $c_j$ than the case in Fig. \ref{fig_f_discrete_1}. For several arbitrary values of $N_{\rm min}(t)$, the figure displays $\sum_{l=l_0-e}^{l_0+e}  P(l)$ as a function of $a$, with $e$ and $a$ satisfying ~\eqref{eq:le2}. The lower bound derived in  \eqref{Prob2bis} is represented as the top of the pink filled area. 
We observe that the lower bound is quite close to the actual probabilities, i.e. it provides a good idea of the probability of success. More importantly, we observe that the probabilities have some erratic behavior w.r.t. $N_{\rm min}(t)$ (especially at $a=0$, due to the erratic behavior of $\lvert f(.) \rvert^2$) but converge quickly towards $\lvert c_0\rvert^2$ as $a$ increases. The usage of $a\ge 1$ thus helps to mitigate the erratic behavior. In particular, we observe that if $N_{\rm min}(t) = 4 $ or $N_{\rm min}(t) = 7$, 
taking $a=2$ helps to increase sharply the probability of success.
However, for $N_{\rm min}(t) = 5 $ or $N_{\rm min}(t) = 6$, $a$ must be increased much more. Indeed, in these cases, a small increase of $a$ makes the probabilities worse. This 
is mostly related to `excited phases' $\theta_{j>0}^{(t)}$
associated with non-negligible $\lvert c_j \rvert^2 $ values ($j=1$ and $2$ as visible in Fig.~\ref{fig_f_discrete_1} \textbf{(f), (g), (h), (i)}), whose probability can be `boosted' compared to the probability of the ground state in some discretization configurations (behavior of $|f(.)|^2$ described in section \ref{sec:f}).
Therefore, we cannot conclude that there is always an advantage in considering $a>0$ together with a window of size $2e+1$ in which the probabilities are summed, even if the behavior w.r.t. $N$ is smoother than when no window is considered. 

\subsection{Analysis of $P(.)$ and number of  shots condition}
\label{sec:shots}

We here describe the impact of $P(l)$ in terms of number of required QPE runs. 
If the number of required QPE shots is often expressed as $O(1/P(l_0))$ in the literature, this quantity represents the average number of measures required to read $l_0$ once, which it is not really representative of the number of measures required to ensure that $l_0$ can be unambiguously deduced from the set containing all measures.
We provide in the following a condition on the number of shots to allow to recover unambiguously an estimate of $l_0$.

We first consider that we would like to estimate on average the number of shots such that $l^*=\arg \max_l P(l)$ (supposed unique and equal to $l_0$) is the most read phase.
We define the independent and identically distributed random variable $X$ in $\{0,1,...,2^N-1\}$ such that $\mathcal{P}(X=l) = P(l)$. $m\in\mathbb{N}^*$ QPE phase measurements or shots provide realizations $X_1, X_2, ... X_m$ of $X$. We  then define the following unbiased estimator of $P(l)$:
\begin{align}
    \mathcal{E}(l,m) = \frac{1}{m}\sum_{k=1}^m \mathbf{1}_{(X_k=l)}
    \quad,\quad
    \lim_{m \to \infty} \mathcal{E}(l,m) =  P(l).
\end{align}
Because QPE is resource consuming, our goal 
is to find the minimum number of QPE shots $\widetilde{m}\in\mathbb{N}^*$ such that $l^*$ is the most represented value within the set of QPE outcomes, i.e. such that 
$\text{argmax}_l {\mathcal{E}(l,\widetilde{m})} = l^*$.
Unfortunately, this relation does not allow us to derive an order of magnitude for $\tilde m$. In practice, we have to find another way to bound the number of shots and approximate $\tilde m$.

Reference \cite{shukla2025practicalquantumphaseestimation} proposes a methodology, which we here adapt to our needs. For $j \in \{0, ..., 2^N-1\}$ and $k \in \{1,...,m\}$, they define the random variables $Y_{k,j}^{(l)} = \mathbf{1}_{(X_k = l)} - \mathbf{1}_{(X_k = j)} \in \{-1, 1\}$ and $Z_j^{(l,m)} = \sum_{k=1}^m Y_{k,j}^{(l)}$. Interestingly, $Z_j^{(l,m)}$ describes the number of times measuring the value $l$ is more than measuring the value $j$. 
Applying the Hoeffding's inequality to $Z_j^{(l,m)}$ yields: $\forall t>0, \,\ \mathcal{P}\left(Z_j^{(l,m)} - \mathbb{E}\left(Z_j^{(l,m)}\right) \leq -t \right) \leq e^{-\frac{t^2}{2m}}$, where $\mathbb{E}(Z_j^{(l,m)}) = m\times\left(P(l)-P(j)\right)$. Taking $t=\mathbb{E}(Z_j^{(l^*,m)})$ (which is strictly positive as we suppose $l^*$ is unique), we obtain:
\begin{equation}\label{eq:prob_bound}
     \mathcal{P}\left(Z_j^{(l^*,m)} \leq 0\right) \leq e^{-m\frac{\left(P(l^*)-P(j)\right)^2}{2}} \leq e^{-m\frac{\Delta_{(l^*)}^2}{2}},
\end{equation}
where $\Delta_{(l^*)}$ is defined through ~\eqref{proj_constr_delta2}. Thus, the probability that,  within a set of $m$ QPE shots outcomes, measured phase values equal to $l^*$ have a lower count than any other measured $j$ values is upper-bounded.
It remains to find the number of shots $m_\epsilon$ such that 
$\epsilon$ upper-bounds $\mathcal{P}(Z_j^{(l^*,m_\epsilon)} \leq 0)$ for all $j$, or equivalently such that $1-\epsilon$ lower-bounds $\mathcal{P}(Z_j^{(l^*,m_\epsilon)} > 0)$ for all $j$, which leads to:
\begin{equation}
     e^{-m_\epsilon\frac{\Delta_{(l^*)}^2}{2}} =\epsilon.
\end{equation}
Thus, an upper-bound for the minimum number of shots $m_\epsilon$ to ensure $l_0$ (supposed equal to $l^*$) has a probability $1-\epsilon$ to be the most measured phase is:
\begin{equation}\label{eq:shots-eps}
     m_\epsilon = \left\lceil-\frac{2\ln \epsilon}{\Delta_{(l_0)}^2}\right\rceil
     \quad,\quad 0<\Delta_{(l_0)}\le 1.
\end{equation}
Taking $\epsilon = 10^{-1}$ and supposing $\Delta_{(l_0)} \approx 1$, which is unrealistic (as it means $P(l_0)\approx 1$ and thus perfect initial state and discretization), leads to $m_\epsilon \le 5$.
Considering $\Delta_{(l_0)} \ge 0.2$, which is more realistic and in line with our previous reasoning and results,
leads to $\Delta_{(l_0)}^2 \ge 0.04$ and $m_\epsilon \le 115$.
This means that, in practice, a hundred of QPE shots might be required even with the kind of initial state satisfying ~\eqref{condqpe1}-\eqref{condqpe2} that we consider here.
\begin{figure}[H]
    \centering
    \includegraphics[width=0.8\linewidth]{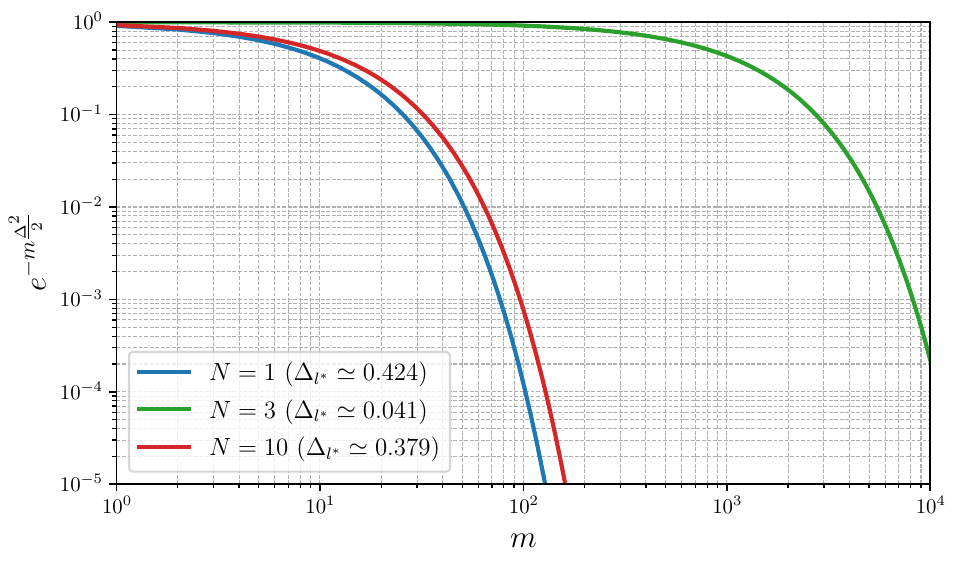}
    \caption{Upper bound of the probability $\epsilon=e^{-m_\epsilon\frac{\Delta_{(l^*)}^2}{2}}$ such that $l^*$ is not the most read phase as a function of the number of shots (taking a sufficiently small value such as $\epsilon=10^{-2}$ is pertinent for our applications). The values of $\theta_j^{(t)}$ and $c_j$ are the same than in Fig. \ref{fig_f_discrete_1}.}
    \label{fig_QPE_nshots}
\end{figure}

Fig. \ref{fig_QPE_nshots} illustrates both the upper-bound probability for several values of $N$ in the same configuration as in Fig.~\ref{fig_f_discrete_1}, as a function of the number of shots $m_\epsilon$. 
According to the $P(l)$ distribution displayed in Fig. \ref{fig_f_discrete_1} and the corresponding $\Delta_{(l^*)}$, 
we see that much less shots are required to reach a given accuracy in the $N=1$ and $10$ cases.
Although the case $N=3$ yields the correct ground phase and is able to project satisfyingly on the ground state, its $\Delta_{(l^*)}$ value is much smaller which requires many more shots to statistically discriminate between $l^*$ measures and other measures with a slightly smaller probability (see the middle panel of Fig. \ref{fig_f_discrete_1}).

Finally, it should be added that it is not straightforward to derive a relation equivalent to ~\eqref{eq:shots-eps} when considering a confidence interval for $l$ values, ~\eqref{Prob2bis}. Indeed, this may require the formalization of a sliding window that takes into account all combinations of nearest neighbors. It should intuitively lead to a $\Delta_{(l_0 \pm e)}$ close to $\lvert c_0\rvert^2 - \max_j \lvert c_j \rvert^2 $ and a smoother behavior of the number of shots  w.r.t $N$, as it leads to a smoother behavior of the probabilities w.r.t $N$, remind the considerations in the end of section \ref{sec:init}.
But it is not straightforward to justify such a $\Delta_{(l_0 \pm e)}$ form within the framework used here, so we keep the investigation for future work.

\subsection{Conclusion on phase qubits}
\label{sec:phase2}

Coming back to the number of phase qubits and QPE ground phase estimation, we highlighted in section \ref{sec:f} the `oscillation' of $|f(\theta_0^{(t)}-\frac{l_0^{(N)}}{2^N})|^2$ in $[0.41,1]$ as $N$ increases, due to discretization effets.
This affects $P(l)$ as highlighted in section \ref{sec:init} and the number of QPE shots $m_\epsilon$  as highlighted in section \ref{sec:shots}.
We noticed how adding just $1$ phase qubits and passing from $N$ to $N+1$ can change the QPE efficiency,
and thus the validity of the conditions in ~\eqref{proj_constr_delta} and~\eqref{proj_constr_2} or~\eqref{proj_constr_8}.
Ideally, we would like the maximum ground phase probability $|f(\theta_0^{(t)}-\frac{l_0^{(N)}}{2^N})|^2$ to be large and above $0.81$ and the $\Lambda_j$ to be small for $j\ge 1$, which depends in a non-trivial way on the chosen number of phase qubits $N$ and is not ensured by choosing $N = N_{\rm min}(t)$, i.e. by the condition developed in ~\eqref{eq:N_constr}.

We suggest a first method to overcome these challenges: launch just few QPE tests (for a given small number of shots) with $N$ equal to $N_{\rm min}(t)+a$ for each $a\in\{0,1,2,3\}$, for instance, and keep for the rest of the process the number of phase qubits that provides the most efficient $l^*$ (i.e. the less variance around the most probable event  for the small number of shots considered).
This method should help to mitigate problems related to discretization (and potentially phase degeneracy), foster a smaller overall number of shots $m_\epsilon$, and lead to a more effective ground state projection (in the conditions explicated above).

We can also leverage the additional qubits $a\in\{0,1,2,3\}$ by the usage of a window of size $2e+1$, with $e$ defined by  \eqref{eq:le2}, in which the occurrence percentage of  the measured phases after $m_\epsilon$ shots are summed.
For instance, taking two additional qubit ($a=2\Rightarrow e=1$) allows to have three phase values that are `chemical precise' and benefit from an enhanced probability of success.
The usage of the window without any prior knowledge can however be challenging to select the QPE phase measures the closest to the ground phase from a set of QPE measures and does not allow optimum control on the ground state projection. It can nevertheless be used as a complement of the first method we proposed above, to confirm the obtained $a$ choice for a given small number of shots considered.

\section{Remarks on the unitary operator approximation, and Trotter steps illustration}
\label{sec:trott}

Given that $H$ naturally takes a LCU form \cite{nielsen2005fermionic, childs2012hamiltonian, loaiza2023reducing}, ~\eqref{eq:LCU}, we have:
\begin{equation}
    U^{2^q} = 
    e^{-iH 2\pi t 2^{q}} = 
    e^{-i\sum_{\beta} H_{\beta} \gamma_\beta 2\pi t 2^q}.
\end{equation}
The conditions developed above for QPE efficiency are exact if the implementation of the $U^{2^{q}}$, denoted by $\mathcal{S}(U^{2^q})$, are exact. 
However, corresponding implementations denoted by $\mathcal{S}(U^{2^q})$ are often approximate and lead to a `perturbed' unitary than can be written (for an `effective' QPE Hamiltonian $H_\mathcal{S}$):
\begin{equation}
\mathcal{S}(U^{2^q})=e^{-iH_\mathcal{S} 2\pi t 2^q}.
\end{equation}
%
We have, as $H$ and $H_\mathcal{S}$ are hermitian 
(see  \cite{childs_theory_2021} corollary A.5):
\begin{align}
    \left\lVert U^{2^q} - \mathcal{S}(U^{2^q})\right\rVert 
    \le  2\pi t 2^{q}  \left\lVert H - H_\mathcal{S}\right\rVert.
\end{align}
Reaching the chemical precision on the 
estimated ground energy requires to first-order (using the spectral norm):
\begin{align}\label{eq:cond_unit}
\left\lVert H - H_\mathcal{S}\right\lVert\lesssim \varepsilon_{\rm ch}.
\end{align}
{We deduce that ~\eqref{eq:cond_unit} implies the following precision condition on the approximate unitary implementation $\mathcal{S}(U^{2^q})$:}
\begin{equation}\label{eq:unit_chem_acc}
    \left\lVert U^{2^q} - \mathcal{S}(U^{2^q}) \right\rVert \lesssim 2\pi t2^q \varepsilon_{\rm ch}=\frac{\pi}{2^{N_\text{min}(t)-q}},
\end{equation}
{the last equality being obtained using the condition in ~\eqref{eq:N_constr}.
A detailed analysis related to this condition go beyond the scope of this article and is kept for a further study.}

\begin{figure}[!h]
    \centering
    \includegraphics[width=0.75\linewidth]{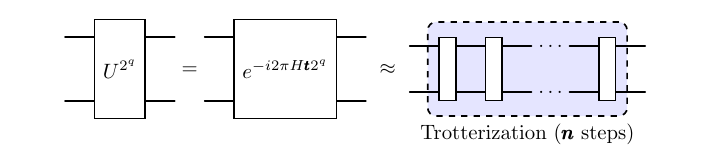}
    \caption{Implementation of $U^{2^q}$ using first-order Trotterization ($p=1$).}
    \label{fig:U2q_implementation}
\end{figure}

We now take the example of the order-$p$ Trotterization approximation \cite{nielsen2010quantum}, denoted by $\mathcal{S}(U^{2^q}, p)$, which represents a common implementation of the QPE unitaries \cite{suzuki_general_1991, rajagopal_generalization_1999}.
E.g., we have $\mathcal{S}(U^{2^q}, 1)=\left( \prod_{\beta} e^{-i H_\beta \gamma_\beta\frac{2\pi t 2^q}{n}} \right)^{n}$, illustrated in Fig. \ref{fig:U2q_implementation}, and the general formulation of $\mathcal{S}(U^{2^q}, p)$ can be found e.g. in  \cite{suzuki_general_1991,rajagopal_generalization_1999}.
In all these formulations, a parameter $n\in\mathbb{N}^*$ that represents the number of Trotter steps must be set. In a QPE context, the number of Trotter steps $n_\text{min}(q,t)$ represents an additional free parameter, which must be adapted for each controlled-unitary ($q$ dependence) and for a given choice of time ($t$ dependency). 

Multiple publications have studied the Trotterization accuracy. Especially,  \cite{childs_theory_2021} bound $\lVert U^{2^q} - \mathcal{S}(U^{2^q}, p) \rVert$. Adapting these works to our case, we have:
\begin{equation}\label{trot_orig}
    \left\lVert U^{2^q} - \mathcal{S}(U^{2^q}, p) \right\rVert \leq C_p\frac{(2\pi t 2^q)^{p+1}}{n(q,t)^p},
\end{equation}
where $C_p$ has the unit of an energy power $(p+1)$ and is built from commutators of the $\gamma_\beta H_{\beta}$
\footnote{
E.g., $C_1 = \frac{1}{2}\sum_{s} \lVert \text{ } [\sum_{r=s+1} \gamma_r H_r, \gamma_s H_s] \text{ } \rVert $.
See  \cite{mehendale_estimating_2025, childs_theory_2021} for details and the general formula of $C_p$.
}.
We deduce from  (\ref{trot_orig}) that an upper-bound for the minimum number of Trotter steps $n_\text{min}(q,t)$ compliant with ~\eqref{eq:unit_chem_acc} should satisfy (using ~\eqref{eq:N_constr}):
\begin{equation}
     C_p\frac{(2\pi t 2^q)^{p+1}}{n_\text{min}(q,t)^p}\approx 
     \frac{\pi}{2^{N_\text{min}(t)-q}}
    \quad\Rightarrow\quad
    n_\text{min}(q,t)\approx 
    \left\lceil\frac{2^q}{2^{N_\text{min}(t)}}\mathscr{C}_p\right\rceil,
\end{equation}
where:
\begin{equation}\label{eq:Cp0}
    \mathscr{C}_p =
    \pi\left(\frac{C_p}{\varepsilon_{\rm ch}^{p+1}}\right)^\frac{1}{p}.
\end{equation}
The number of Trotter steps to reach the chemical precision with a given QPE unitary should thus satisfy:
\begin{equation}\label{trot_tot_steps_unit}
    n(q,t) \gtrsim n_\text{min}(q,t).
\end{equation}
Considering  
$a \ll N_{\rm min}(t)$ and taking the notations of  \eqref{eq:N_constr_more}, the overall minimum number of Trotter steps over the whole QPE circuit is:
\begin{align}\label{eq:Cp}
      n_\text{min-tot}=\sum_{q=0}^{N_\text{min}(t)+a-1}n_\text{min}(q,t) 
      \approx 
      \left\lceil2^a\sum_{q=0}^{N_\text{min}(t)+a-1} \frac{2^q}{2^{N_\text{min}(t)+a}}\mathscr{C}_p\ \right\rceil 
      \approx\left\lceil 2^a \mathscr{C}_p \right\rceil,
\end{align}

where the first $\approx$ is due to the fact that $\frac{1}{2^{N_\text{min}(t)}}\mathscr{C}_p$ is much larger than $1$ for many-electron systems (which limits ceiling effects), and the second $\approx$ is obtained using $\sum_{q=0}^{N-1}\frac{2^q}{2^{N}}=1-\frac{1}{2^{N}}\approx 1$ for reasonably large $N$ (here meaning $N\gtrsim 6$)
\footnote{
Note that the result in  \eqref{eq:Cp} could equivalently be obtained by constraining the spectral norm related to the whole QPE (product $\prod_{q=0}^{N_\text{min}(t)+a-1}$ of unitaries), which would lead to replace the $2^q$ in  \eqref{eq:cond_unit} by $\sum_{q=0}^{N_\text{min}(t)+a-1}2^q$. This means that constraining the chemical precision for each of the unitaries,  (\ref{eq:cond_unit}), is equivalent to constraining the chemical precision for the whole circuit.
}.

Interestingly, $n_\text{min-tot}$ can be considered as mostly dependent on the physical system features (more specifically the LCU decomposition of $H$) and independent of $N_\text{min}(t)$ and thus of $t$. This represents a property of the application of QPE to many-electron systems. A dependence on the number of added phase qubits $a$ (beyond the chemical precision requirements) remains in the final result, but this dependence is weak (less than an order of magnitude if $a \leq 3$).

{
$\mathscr{C}_p$, ~\eqref{eq:Cp0}, thus directly gives a measure of the overall number of Trotter steps required by the controlled-unitaries.
To give an order of magnitude,
in the case of the \chemform{H_2} molecule with sto-3g basis set and $0.5$ $\rm{\AA}$ bond length, we approximately have $C_1 \simeq 0.052$ and $C_2 \simeq 0.055$, and thus $\left\lceil \mathscr{C}_{1} \right\rceil \simeq 6\times 10^4$ an $\left\lceil \mathscr{C}_{2} \right\rceil \simeq 1 \times 10^4$, which implies less that an order of magnitude difference in total steps between first and second-order Trotterization. 
More generally,  \cite{blunt_monte_2025} gives a methodology for estimating $C_p$ and thus $\mathscr{C}_{p}$ in the general case with a Monte-Carlo approach.
}

{
The complexity related to the Trotterized controlled-unitaries thus equals 
\begin{align}\label{eq:T_complexity}
 {\lceil 2^a \mathscr{C}_p\rceil} \times \mathscr{N}_p(N_S),
\end{align}
where $\mathscr{N}_p(N_S)$ represents the complexity related to one order-$p$ Trotter step (measured for instance in terms of number of non-Clifford gates), polynomial in $N_S$ \cite{PRXQuantum.4.020323}.
}

Finally, note that the upper-bounds we have used are known to be quite loose \cite{mehendale_estimating_2025}, as will be illustrated in the next section (an order of magnitude below these bounds will reveal sufficient on a \chemform{H_2} molecule test). Also, we underline that standard order-$p$ Trotterization was taken as an illustration of the consequences of ~\eqref{eq:unit_chem_acc} but more precise Trotterizations and refined bounds have been developed in the literature, e.g. see the recent works in  \cite{Mart_nez_Mart_nez_2023,ikeda_measuring_2024,zhuk_trotter_2024,daspal_minimizing_2024,mehendale_estimating_2025,blunt_monte_2025}. Also, even if Trotterization represents a common implementation of the QPU unitaries, it is not the only possibility \cite{babbush2018encoding, low2019hamiltonian, loaiza2023reducing}. An extensive study of the QPE unitaries implementation goes beyond the scope of this article.

\section{First illustrations on \chemform{H_2}}
\label{sec:4}

\begin{table}[H]
    \caption{\centering Initial data for \chemform{H_2}.}
    \centering
    $\begin{array}{|r|l||c|c|c|c|c|}
         \hline \multicolumn{2}{|r||}{E_{\rm init}}  & \multicolumn{5}{l|}{ -1.042996}  \\
         \hline \multicolumn{2}{|r||}{E_0}  & \multicolumn{5}{l|}{-1.055160}  \\
         \hline\hline \multicolumn{2}{|c||}{t} & \lceil E_0 t \rceil & \lceil E_{\rm init} t \rceil & N_{\rm min}(t) & n_\text{min}(0,t) & n_\text{min-tot} (a=0)\\
         \hline 10^{d} & 10 & -10 & -10 & 5 & 1875 & 6\times 10^4 \\
         \hline -\frac{3}{2}\frac{1}{E_{\rm init}} & 0.713827 & -1 & -1 & 9 & 118 & 6\times 10^4\\
         \hline \frac{1}{2\sum_\beta \lvert\gamma_\beta\rvert} & 0.215149 & 0 & 0 & 11 & 30 & 6\times 10^4\\
         \hline
    \end{array}$
    \label{tab:intial_data_H2}
\end{table}

We illustrate the QPE conditions and features described above on the \chemform{H_2} molecule with $0.5$ $\rm{\AA}$ bond length. We work with the sto-3g basis, where each \chemform{H} is represented with a 1s orbital, leading to a $N_S=4$ qubits system. Note that we are interested in chemical precision and success probability features with respect to the chosen orbital basis.
Succeeding here is a necessary condition to reach chemical accuracy,
which would require to consider much larger orbital sets.
Exact eigen-energies $E_j$ are obtained by a full diagonalization of the Hamiltonian, which also gives the exact eigen-states $\ket{\psi_j}$. The initial state $\ket{\psi_{\rm init}}$ is obtained by a Hartree-Fock computation, and is already of good quality as visible in Table \ref{tab:intial_data_H2}. But it does not allow us to reach the chemical precision and is thus is perfectible as we will highlight below. 
All calculations were performed on the Quantum Learning Machine (QLM) from Bull, which enables large-scale emulations of quantum processing units using the myQLM package. 
The QPE version studied in this section was implemented using a first-order Trotterization.
The values of $t$ derived in section \ref{sec:time} together with the corresponding minimum number of phase qubits to reach chemical precision and Trotter steps are presented in Table \ref{tab:intial_data_H2}.

\begin{figure}[H]
    \centering
    \includegraphics[scale=0.65]{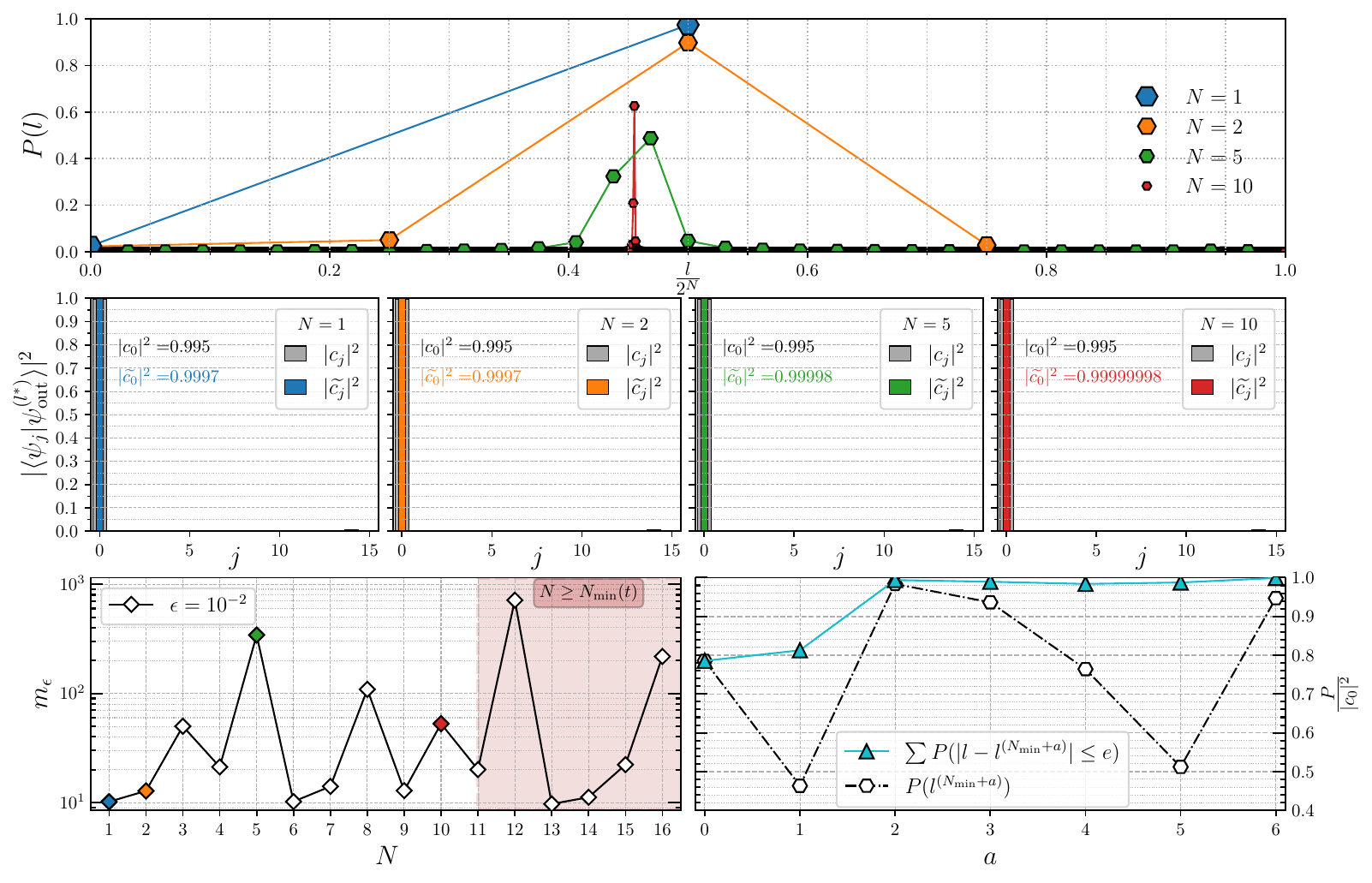}
    \put(-250, 320){(a)}
    \put(-420, 218){(b)}
    \put(-310, 218){(c)}
    \put(-200, 218){(d)}
    \put(-90, 218){(e)}
    \put(-365, 114){(f)}
    \put(-145, 114){(g)}
    \caption{QPE features on \chemform{H_2} with $t=\frac{1}{2\sum_\beta \lvert\gamma_\beta\rvert}$  choice.
    \textbf{(a)} $P(l)$ as a function of $\frac{l}{2^N}$ for several $N$ values. \textbf{(b), (c), (d), (e)} Overlap between each exact eigenstate of $H$ with the initial state ($|c_j|^2$, grey) and the final state after phase measurement ($|c_j^{(l^*)}|^2$, color), for several  $N$ values. \textbf{(f)} Upper bound of the minimum number of shots $m_\epsilon$ required so that $l^*$ is the most read phase with $1 - \epsilon$ probability (here $\epsilon = 10^{-2}$). \textbf{(g)} Probability $P(l_0)$ of the ground phase compared to the probability $\sum_{l=l_0-e}^{l_0+e}  P(l)$ associated to the case in ~\eqref{e} and \eqref{everyfirst}, for various values of the additional number of phase qubits $a$ (beyond chemical precision).}
    \label{fig_H2_f_discrete}
\end{figure}

Fig.~\ref{fig_H2_f_discrete} shows the same kind of QPE features as in Figs.~\ref{fig_f_discrete_1} and \ref{fig_P_relax_minor}. However, the $\theta_j^{(t)}$ are now the exact eigen-phases related to the choice $t=\frac{1}{2\sum_\beta \lvert\gamma_\beta\rvert}$. The $\lvert c_j \rvert^2$ related to the initial Hartree-Fock state $\ket{\psi_{\rm init}}$ can be computed using the $\ket{\psi_j}$. We observe on \textbf{(b), (c), (d)}, and \textbf{(e)} that $\lvert c_0\rvert^2$ is extremely dominant and very close to $1$, confirming that the Hartree-Fock state is already very good.
On Fig.~\ref{fig_H2_f_discrete} \textbf{(a)} we observe a unique peak for $N=1, 2$ and $10$, which is due to the fact that $|c_0|^2$ is large (as the Hartree-Fock state is already very good). However, in the case $N=5$, we observe two smaller peaks close to $0.5$, which denotes a strong discretization effect ($\theta_0^{(t)}$ lies almost in-between a discretization interval with $\kappa_0^{(N=10)}\approx 0.2$). This leads to a quite large shot number to be able to recover the ground phase $l^*$, as visible from \textbf{(f)} pannel. While having a peak higher than $0.6$, the case $N=10$ also has a second non-negligible peak which tends to increase the number of shots required. In both $N=5$ and $10$ cases, we observe however that adding only $1$ qubit ($N=6$ and $11$) allows us to strongly improve the situation, visible from the shot number point of view in \textbf{(f)}.
The shot number `plateau' for $N \in \{13,15\}$ indicates that the situation remains good for these number of qubits (the ground phase probability remains large), but suddenly becomes bad fo $N=16$. This justifies to use the refined method we proposed in section \ref{sec:phase2} starting from $N_{\rm min}(t) = 11$ (minimum number of phase qubits to reach the chemical precision for the $t=\frac{1}{2\sum_\beta \lvert\gamma_\beta\rvert}$ choice considered in the figure). 
Fig.~\ref{fig_H2_f_discrete} \textbf{(g)} shows that considering $N_{\rm min}(t)+a$ qubits can be interesting to increase the probability of success of QPE. The behavior of $P(l_0)$  w.r.t. $a$ is erratic as awaited, whereas the behavior of $\sum_{l=l_0-e}^{l_0+e}  P(l)$, ~\eqref{e} and \eqref{everyfirst}, increases regularly and becomes close to $1$ for $a \geq 2$ thanks to a favorable $\lvert c_0 \rvert$ condition. 
Consequently, going from $N = N_{\rm min}(t)$ to $N_{\rm min}(t) +2$ phase qubits allows to improve the efficiency of QPE whatever the option considered.
Regarding ground state projection (in Fig. \ref{fig_H2_f_discrete} \textbf{(b), (c), (d), (e)}), the consequence of having $\lvert c_0\rvert^2$ already close to $1$ is to make its amplification less visible. However, an amplification occurs for $N=5$ and $10$ as visible from the numbers printed in the middle figures. 

\begin{figure}[H]
    \centering
    \includegraphics[width=\linewidth]{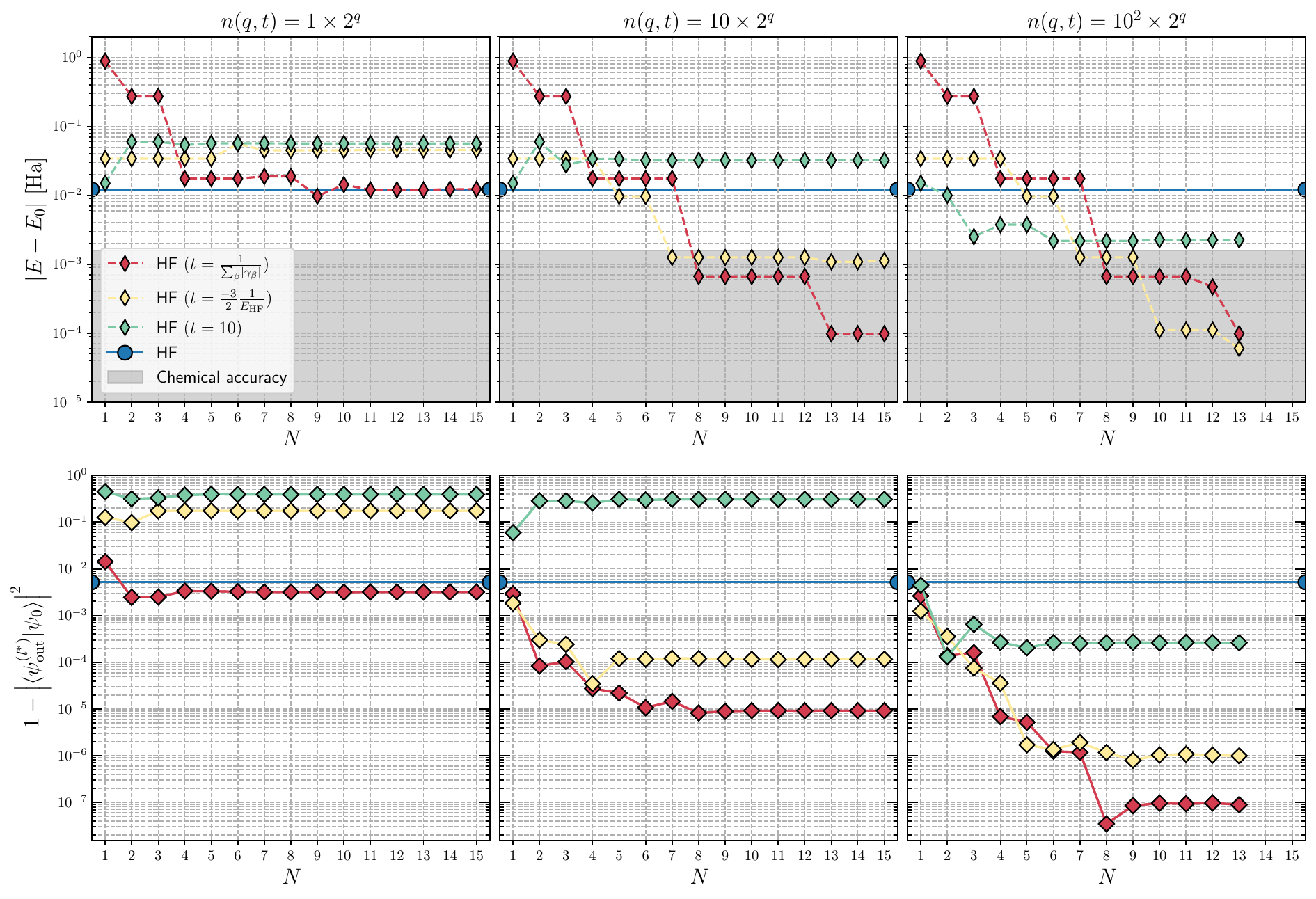}
    \put(-480, 342){(a)}
    \put(-320, 342){(b)}
    \put(-160, 342){(c)}
    \put(-480, 170){(d)}
    \put(-320, 170){(e)}
    \put(-160, 170){(f)}
    \caption{QPE result on \chemform{H_2}. The three values of $t$ presented in Table \ref{tab:intial_data_H2} are benchmarked (corresponding to the lines with colored diamond markers), as well as the number of Trotterization steps (equal to $n_\text{min}(q,t) = 1\times 2^q$, $10\times 2^q$ and $100\times 2^q$). 
    \textbf{(a), (b), (c)} Difference between the energy reconstructed with QPE and the exact ground energy of the \chemform{H_2} system, as a function of the number of phase qubits. The Hartree-Fock energy (more exactly $\lvert E_{\rm init} - E_0 \rvert$) is represented with a horizontal line with circles at the ends. The chemical precision region (w.r.t $E_0$) is represented as a gray span.
    \textbf{(d), (e), (f)} Fidelity of the QPE projected state after measurement of the most probable phase with respect to the exact ground state, as a function of the number of phase qubits. The horizontal blue line denotes the overlap with the initial Hartree-Fock state.}
    \label{fig_QPE_H2_as_ntr}
\end{figure}

Fig. \ref{fig_QPE_H2_as_ntr} \textbf{(a), (b), (c)} focus on ground energy estimation using QPE. It shows the difference between the energy reconstructed with QPE and the exact ground energy of the \chemform{H_2} system, as a function of the number of phase qubits on $x$-axis. The three values of $t$ presented in Table \ref{tab:intial_data_H2} are benchmarked (corresponding to the lines with colored diamond markers), as well as the number of Trotterization steps $n_\text{min}(q,t)$, equal to $1\times 2^q$ in the left figure, $10\times 2^q$ in the middle figure and $100\times 2^q$ in the right figure,
{
the factors in front of $2^q$ ($1$, $10$ and $100$) being to be compared to the $n_\text{min}(0,t)$ in Table \ref{tab:intial_data_H2} to understand which value is closer to the a theoretical optimum.
} 
The initial Hartree-Fock energy $E_{\rm init}$ is represented with a horizontal line with circles at the ends. For reference, the chemical precision region (w.r.t $E_0$) is represented as a gray span. 
One can notice that, even if there is an overlap of $99.5 \%$ of the Hartree-Fock state w.r.t exact ground state,
the Hartree-Fock energy is outside of the chemical precision region (approximate $10^{-2}$ error in Hartree). 
This is what we want to improve using QPE.
The left figure shows that $n_\text{min}(q,t)=1\times 2^q$ is not sufficient to improve the initial guess. 
Improving the Trotterization steps by an order of magnitude (middle figure) leads to a convergence under chemical precision for $t=-\frac{3}{2}\frac{1}{E_{\rm init}}$ when $N \geq 7$ and $t=\frac{1}{2\sum_\beta \lvert\gamma_\beta\rvert}$ when $N \geq 8$,
so with lower values than the $N_{\rm min}(t)$ values in Table \ref{tab:intial_data_H2}. 
Indeed, $N_{\rm min}(t)$ corresponds to the value of $N$ for which it is certain that the discretization error is below the chemical precision condition, but it does not prevent in specific situations to reach the chemical precision with less qubits. 
Now, when $N\geq N_{\rm min}(t)$, the residual error is only induced by the first-order Trotterization, and converging even if the inequalities in ~\eqref{trot_tot_steps_unit}-\eqref{eq:Cp} are not satisfied confirms that corresponding upper-bounds (see Table \ref{tab:intial_data_H2}) are quite loose. 
Increasing the number of Trotter steps by an additional order of magnitude in \textbf{(c)} makes this number close to the upper-bounds in ~\eqref{trot_tot_steps_unit}-\eqref{eq:Cp} or Table \ref{tab:intial_data_H2}. We then observe a similar behavior between the yellow and red curves. The green curve ($t=10$) approaches the chemical precision region without converging. 
Consequently, the optimal $t$ for determining the ground energy with QPE might here be $t=-\frac{3}{2}\frac{1}{E_{\rm init}}$ or $t=\frac{1}{2\sum_\beta \lvert\gamma_\beta\rvert}$, given that both allows us to converge, are different from just $1$ phase qubit and require similar number of Trotter steps. Interestingly, $N_{\rm min}(t) = 9$ for $t=-\frac{3}{2}\frac{1}{E_{\rm init}}$ (see Table \ref{tab:intial_data_H2}), which should allow for a small number of shots (see Fig. \ref{fig_H2_f_discrete} \textbf{(f)}).

Fig. \ref{fig_QPE_H2_as_ntr} \textbf{(d), (e), (f)} focus on ground state projection using QPE. It analyses the system register state after the most probable phase measurement. We calculated the overlap of this state with the exact ground state as a function of the number of phase qubits. The horizontal blue line denotes the overlap with the initial Hartree-Fock state. 
We notice that the evolution of the overlap with $N$ is much smoother than the energy convergence in Fig. \ref{fig_QPE_H2_as_ntr}. Surprisingly, \textbf{(d)} shows that $t=\frac{1}{2\sum_\beta \lvert\gamma_\beta\rvert}$ is able to improve the quality of the initial Hartree-Fock state (for $N \geq 2$) while the ground energy was not more precise than $E_{\rm init}$. This highlights the lack of correlation between state overlap and energy accuracy. 
Significant improvement of the initial state can be observed in \textbf{(e)} and \textbf{(f)} figures. Notably, when the  condition $N \geq N_{\rm min}(t)$ is met, the projection does not improve for any of the three values of $t$. Moreover, the overlap values are not necessarily matching the forecasted coefficients.
In \textbf{(e)} and \textbf{(f)}, we observe that, for $n_\text{min}(q,t) = 10\times 2^q$ and $n_\text{min}(q,t) = 100\times 2^q$, the results tend to improve with more phase qubits and more Trotter steps. Even in situations where the Hartree-Fock state is already very good, it can be improved by QPE, which is encouraging.
Overall, these results highlight the importance of the quality of implementation of the unitaries. Here Trotterization has a non-trivial impact, which can be advantageous or disadvantageous. 
Also, the optimal $t$ for projection might be $t=\frac{1}{2\sum_\beta \lvert\gamma_\beta\rvert}$, given that it allows us to more strongly improve the initial state even with suboptimal Trotterization condition. 

\section{Conclusion}
\label{sec:conclusion}

We presented an analysis of QPE applied to many-electron systems. We reminded that such an application, while theoretically powerful, requires careful tuning of free parameters to deliver precise results for ground energy estimation and ground state projection. 
{We gathered and refined constructive conditions for time step, number of phase qubits and number of shots. We analyzed initial state requirements and emphasized the impact of the approximations done when implementing the QPE unitary operators. We highlighted that the complexity of the Trotterized version of QPE tends to depend {mostly} on physical system features and {weakly} on the number of phase qubits.
We showed how these factors collectively govern QPE efficiency in terms of probability of success and precision.
We detailed the strong impact of the QPE `blurring function', related to discretization effects, and proposed a new method to overcome corresponding pathologies.
Various numerical results illustrated the impact of our points on the success probability and discretization effects.
A numerical experiments on \chemform{H_2} illustrated that parameter optimization can significantly impact QPE precision on both ground energy estimation and ground state projection, and allowed us to derive useful insights.
Overall, this study offers practical and standalone guidance towards automation of QPE for predictive computational chemistry and material science applications, with optimized resources that enable a controlled accuracy.
}

The constructive conditions {above} are general to many-electron systems and we plan to analyze their behavior on molecular systems larger than \chemform{H_2},
as well as the practical advantages of using more phase qubits than required by the chemical precision condition to increase the probability of success of QPE.
It would also be interesting to conduct statistical analysis on the values of $\lvert c_0\rvert^2$ commonly given by Hartree-Fock or CI states, for a controlled database of chemical systems. 
The preparation of a sufficiently good initial state (i.e. that matches the conditions detailed in this article) seems to be the most important challenge of QPE.
Indeed, the rapid decay with $N_S$ of the overlap with the true ground state for systems that are difficult to describe classically seems to suggest that practical uses of QPE may remain confined to intermediate‑size molecules, thus to the non‑asymptotic regime \cite{lee2023evaluating,louvet2023feasibility}.
A study of the complexity $\mathscr{N}_p(N_S)$ required by an order‑$p$ Trotterization step (measured for instance in terms of number of non-Clifford gates on real quantum processing units) including prefactors would probably be crucial.
Further work should include decomposition in elementary gates, optimized among others by leveraging properties of the considered many-electron system
to reduce as much as possible the degree of the polynomial in $N_S$ and the prefactor. 
This would help us
refine roadmaps for QPE utilization on industrially interesting cases.

{
Finally, as the size $N_S$ of the system increases, more and more eigenvalues (at most $2^{N_S}$ different values) are mapped into the interval $[0,1[$ as QPE works with phases, equation~\eqref{E_j}. So, investigating statistically the corresponding distribution of phases (to understand the potential reinforcement of probabilities associated to `excited phases' due to overlapping) could allow to  refine the proposed constructive method use on larger systems. Moreover, the impact of $t$ and $a$ parameters on this degeneracy remains to be studied.
}
We also mention the potential pertinence of elements of our analysis in the context of linear systems resolution \cite{HHL2009}.

\section*{Acknowledgment}
We would like to warmly thank Coretin Bertrand (Bull), Jean-Patrick Mascom\`ere, Henri Calandra, Yagnik Chatterjee (TotalEnergies) and Matthieu Saubanère (LOMA, CNRS) for their feedback and insightful comments on the present article. 
We thank TotalEnergies for the permission to publish this work.

\bibliography{bibli}

\EOD

\end{document}